\documentclass[
    reprint,
    superscriptaddress,
    nofootinbib,   
    amsmath,amssymb,
    aps,
    prc,floatfix
]{revtex4-2}
\usepackage{style}

\begin{document}

\preprint{APS/123-QED}

\title{Nuclear Data Evaluation Utilizing Bayesian Monte Carlo Methods}

\author{Catherine E. Apgar}
\email{ceapgar@lanl.gov}
 \affiliation{Los Alamos National Laboratory, Los Alamos, New Mexico 87544, USA}
 
\author{Lee A. Bernstein}
\affiliation{Department of Nuclear Engineering, University of California, Berkeley, Berkeley, California 94720, USA}

\affiliation{Lawrence Berkeley National Laboratory, Berkeley, California 94720, USA}
\author{Yun-Hsuan Lee}
\affiliation{Department of Nuclear Engineering, University of California, Berkeley, Berkeley, California 94720, USA}

 \author{Wilson Lin}
 \affiliation{Brookhaven National Laboratory, Upton, New York 11973, USA}
 
\author{Alan B. McIntosh}
\affiliation{Texas A\&M University Cyclotron Institute, College Station, Texas 77843, USA}

 \author{Dmitri G. Medvedev}
 \affiliation{Brookhaven National Laboratory, Upton, New York 11973, USA}

 \author{Jonathan T. Morrell}
 \affiliation{Brookhaven National Laboratory, Upton, New York 11973, USA}

  \author{Ellen M. O'Brien}
 \affiliation{Los Alamos National Laboratory, Los Alamos, New Mexico 87544, USA}

 \author{C. Etienne Vermeulen}
 \affiliation{Los Alamos National Laboratory, Los Alamos, New Mexico 87544, USA}
 
\author{Andrew S. Voyles}
\affiliation{Department of Nuclear Engineering, University of California, Berkeley, Berkeley, California 94720, USA}

\date{\today}

\begin{abstract}
    While the body of experimental nuclear data continues to grow at a rapid pace, significant gaps persist.
The lack of evaluated experimental data is far more pronounced in charged-particle reactions which are essential to a large portion of the isotope production community for vital medical, industrial, and national security interests.
In the absence of experimental and evaluated data, isotope production relies on theoretical models that may result in low yield and isotopic impurities. 
This work investigates model and parameter adjustments in TALYS 2.0 to establish uncertainty bands on proton-induced cross sections, highlight influential parameters in calculations, and constrain adjustments within physically defensible bounds.
The EXFOR dataset for proton-induced reactions on natural copper, comprised of 24 independent cross sections, was used for iterative Bayesian model updating. 
Upon convergence, parameters and their uncertainties were compared against 23 cumulative cross sections from EXFOR, highlighting the limitations of generalization.
Various machine learning models were leveraged to predict TALYS performance, with a small neural network showing useful ranking capabilities across several generations, suggesting the use of a surrogate approach to screening for future evaluations.

\end{abstract}

\maketitle

    \section{\label{sec:Introduction}Introduction\protect}
        The nuclear data pipeline is a rigorous, expensive and time-intensive process, from identification of nuclear data needs through to publication and application.
Evaluated nuclear data libraries, such as ENDF/B-VIII.0 \cite{brown_endfb-viii0_2018}, are compiled using experimental measurements and extensive evaluation by subject matter experts in the nuclear data field.
Despite these robust libraries' data for neutron-induced reactions, evaluated data remain sparse for charged particle reactions of interest to the isotope production community. 
Many of these channels have not been sufficiently measured, necessitating the use of theoretical models for cross section predictions.
The nuclear reaction modeling code, TALYS-2.0 \cite{koning_talys_2023}, has been leveraged for this purpose, as seen in a steadily growing number of citations, and with the broad adoption of the TALYS-Evaluated Nuclear Data Library, TENDL-2023 \cite{koning_tendl_2019}.

TENDL provides evaluated cross sections using a combination of experimental nuclear data and randomized parameter variation established by the user to establish a goodness of fit through TASMAN-2.0\cite{koning_tendl_2019}.
The parameters in an individual theoretical calculation can be systematically varied on a per-keyword basis, or globally in TASMAN, to calculate a probability distribution based on the outputs from TALYS. 
The flexibility in TALYS' over 400 tunable parameters provides users with options to explore variations to nuclear reaction models, and these options grow exponentially when individual parameters can be further adjusted for each residual product.
This establishes an optimization procedure using Bayesian Monte Carlo (BMC) methods to assign a weight for each set of TALYS input parameters and an optimized parameter set with an uncertainty band.
This research explores alternatives to the current method by adjusting the random sampling window to a lognormal distribution around the default value utilizing Sobol sequencing\cite{sobol_distribution_1967} for optimal coverage within physically defensible boundaries as well as different methods for the $\chi^2$ goodness of fit to experimental data.
With hundreds of knobs to fine-tune within TALYS, we focus on 85 parameters, similar to Alhassan \etal \cite{alhassan_iterative_2022} to explore their predictive capabilities within TALYS when compared to experimental data. 
We explore various methods to establish a likelihood function and weighting factor for each iteration as well as normalization to reduce overfitting.
This work additionally explores a proposal-posterior approach to develop a weighted ensemble and refine the iterative process. 
The results of the iterative process are calculated using individual residual cross section measurements and validated against cumulative residual cross sections to compare a holistic approach against a refined, individual residual product approach. 
While these results emphasize the continued need for a physics-informed evaluation, they also leave room for supplementary screening via neural networks to potentially down-select during the process to reduce compute and resources. 
The research proposed here is not intended to reproduce or improve on evaluated nuclear data efforts, but rather to explore alternatives and recognize their strengths and weaknesses.

    \section{\label{sec:Background}Background\protect}
        A variety of Bayesian Monte Carlo methods have been used in the evaluation of nuclear data and theoretical modeling of reactions\cite{koning_bayesian_2015,koning_modern_2012,capote_ripl_2009,neudecker_machine_2025,neudecker_advanced_2014,neudecker_informing_2021}.
This analysis establishes both predictions and their associated uncertainties for calculated cross sections, both for reactions that have been experimentally measured and for reactions for which sparse data exists. 
Experimental data is costly, from facility costs to analysis, publication, and evaluation. 
It is therefore imperative to have more robust theoretical models. 
The majority of evaluation effort has historically focused on neutron-induced reactions and benchmark evaluations with sparse data for charged particle reactions valuable for the isotope production community\cite{capote_iaea_2017}.
Evaluated charged particle reaction datasets are sparse, necessitating the use of theoretical models. 
The isotope production community relies on these models to inform production routes. 
TALYS, and the associated evaluated nuclear data library TENDL, have been leveraged for over 20 years to predict cross section measurements, and consequently, theoretical yields.
TENDL is a trusted source for theoretical calculations and has been developed using an iterative Monte Carlo method through TASMAN and related T6 software\cite{koning_talys_2023}. 
This is a successful interpretation of neutron-induced reactions, but is focused on neutron-induced experimental data.

Sparse datasets necessitate theoretical models, as well as associated uncertainties.
One approach is the use of a Unified Monte Carlo method, which has been used for over a decade for analysis.\cite{rochman_tendl_2017,trkov_endf-6_2018}

The approach in this paper seeks to better quantify correlations not only among parameter adjustments in TALYS, but also among residual product results.
The experimental dataset for this analysis uses proton-induced reactions on natural copper (\ce{^{nat}Cu}(p,x)) residual product production, as this is currently the largest dataset in the experimental nuclear data library, EXFOR, with over 2600 independent cross section residual product data points. 
We verify against the cumulative residual product cross sections from proton-induced reactions on \ce{^{nat}Cu}.

In a broader overview, a target nucleus, bombarded by a proton, can produce hundreds of residual products with enough incident particle energy. 

While these cross sections are characterized, potential contaminants, including stable, long-lived, or otherwise unmeasurable residual products are important to characterize, motivating the use of theoretical models.

    \section{\label{sec:Methods}Methods\protect}
        \subsection{\label{subsec:expdata}Experimental Data Selection}

The focus of this paper is to explore the data space of residual product cross sections for proton-induced reactions in the energy region up to 200\,MeV.
Here, we assume cross sections labeled as independent, (\textit{i}) are directly produced from the reaction, and those measured as cumulative, (\textit{c}) include production of parent residual products.
In selecting the ideal target material for this study, several factors were considered.
Natural copper (\ce{^{nat}Cu}) was selected as the target of interest for its abundance of experimental residual product cross section data from 34 distinct experiments, totaling roughly 2600 independent data points and 1600 cumulative data points. 
This analysis utilized 24 independent residual product cross sections for iteration, with the final iteration validated using data from 23 cumulative cross sections.
These isotopes are listed in \autoref{tab:ind,cum}.

\begin{table}
\caption{Residual product cross sections used}
\label{tab:ind,cum}
\begin{tabular}{lll}
\hline
\ce{^{44}Sc_{(i,c)}} & \ce{^{54}Mn_{(i)}}     & \ce{^{61}Co_{(i,c)}} \\
\ce{^{44m}Sc_{(i)}}  & \ce{^{56}Mn_{(c)}}     & \ce{^{56}Ni_{(c)}}   \\
\ce{^{46}Sc_{(i)}}   & \ce{^{52}Fe_{(c)}}     & \ce{^{57}Ni_{(i,c)}} \\
\ce{^{47}Sc_{(c)}}   & \ce{^{52g}Fe_{(c)}}    & \ce{^{60}Cu_{(i,c)}} \\
\ce{^{48}Sc_{(i,c)}} & \ce{^{59}Fe_{(i,c)}}   & \ce{^{61}Cu_{(i,c)}} \\
\ce{^{48}V_{(c)}}    & \ce{^{55}Co_{(c)}}     & \ce{^{64}Cu_{(i)}}   \\
\ce{^{48}Cr_{(c)}}   & \ce{^{56}Co_{(i,c)}}   & \ce{^{60}Zn_{(i)}}   \\
\ce{^{49}Cr_{(c)}}   & \ce{^{57}Co_{(i,c)}}   & \ce{^{61}Zn_{(i)}}   \\
\ce{^{51}Cr_{(c)}}   & \ce{^{58}Co_{(i)}}     & \ce{^{62}Zn_{(i)}}   \\
\ce{^{52}Mn_{(c)}}   & \ce{^{58g}Co_{(i)}}    & \ce{^{63}Zn_{(i)}}   \\
\ce{^{52g}Mn_{(i,c)}}& \ce{^{58m}Co_{(i)}}    & \ce{^{65}Zn_{(i)}}   \\
\ce{^{52m}Mn_{(i,c)}}& \ce{^{60}Co_{(i,c)}}   &                      \\
\hline
\end{tabular}
\end{table}
Several methods were considered for the handling of experimental data outliers.
Previous methods used a binary approach to entirely reject an experiment's datasets when these datasets were observed to be inconsistent when compared to other experiments. \cite{alhassan_bayesian_2020,alhassan_iterative_2022}
This is at the discretion of the evaluator, which inherently introduces bias in the evaluation, however this is an accepted process with other evaluated nuclear data libraries.
In addition, datasets with no listed uncertainties, or uncertainties less than 5\%, were assigned a 15\% uncertainty. 
This increase in the uncertainty of experimental data reduced the very large $\chi^2$ values that could be potentially introduced by inaccurate data subsequently leading to overfitting on these data points.

\subsection{Selection of TALYS Models and Parameters}
This analysis leveraged the TALYS nuclear reaction modeling code. 
One of the most widely used models, TALYS, has been cited or used as a comparator in over 4,000 publications\cite{koning_talys_2023}.
The TALYS modeling code is robust, with  nearly 400 tunable inputs for adjusting theoretical models and their associated parameters. 
It is not computationally feasible to cover this full range of inputs, nor would such an exhaustive search necessarily yield improvement in theoretical fit as a function of the required compute.

There are a few different input types available in TALYS. 
Discrete model adjustments invoke different models or modifications to the default model. 
An example of this is \texttt{preeqmode}, which allows the user to select one of four models for pre-equilibrium reactions, seen in \autoref{tab:preeqmode}.
\begin{table}
\label{tab:preeqmode}
\caption{Pre-equilibrium model options available in TALYS. \texttt{preeqmode}=4 was excluded from this work due to both rigorous compute and non-physical results.}
\begin{tabular}{cc}
\hline \hline
\texttt{preeqmode} & Description \\
\hline
1 & Exciton model: Analytical transition rates \\
 & with energy-dependent matrix element. \\
2 & Exciton model: Numerical transition rates \\
 & with energy-dependent matrix element. \\
3 & Exciton model: Numerical transition rates \\
 & with optical model for collision probability. \\
4 & Multi-step direct/compound model. \\
\hline
\end{tabular}
\end{table}

Adjusting this, and other discrete model adjustments is at the user's discretion. Aside from some hard-coded limitations, these can be combined in any assortment. 
With compute limitations, a subset of these inputs was selected, as seen in \autoref{tab:discrete_params}.


\begin{table}[htbp]
\centering
\caption{Discrete model parameters sampled in Generation 0.}
\label{tab:discrete_params}
\footnotesize
\begin{tabular}{cccc}
\hline \hline
Parameter & Values & Count & Description \Tstrut \Bstrut \\
\hline
\texttt{ldmodel}         & 1--6     & 6  & Level density model \Tstrut\\
\texttt{strength}        & 1--10    & 10 & GSF\\
\texttt{spincutoffmodel} & 1--2     & 2  & Spin cutoff model\\
\texttt{mpreeqmode}      & 1--2     & 2  & Multiple pre-equilibrium mode \\
\texttt{preeqmode}       & 1--3     & 3  & Pre-equilibrium emission model \\
\texttt{preeqspin}       & 1--3     & 3  & Pre-equilibrium spin dist. \\
\texttt{ldmodelracap}    & 1--3     & 3  & Level density - radiative capture \\
\texttt{parity}          & y/n      & 2  & Parity distribution \\
\texttt{massmodel}       & 0--3     & 4  & Nuclear mass model \\
\texttt{gshell}          & y/n      & 2  & Ground-state shell effects \\
\texttt{shellmodel}      & 1--2     & 2  & Shell model correction \\
\texttt{kvibmodel}       & 1--2     & 2  & Vibrational enhancement model \\
\texttt{pairmodel}       & 1--2     & 2  & Pairing correction model \\
\texttt{phmodel}         & 1--2     & 2  & Particle-hole state density \\
\texttt{strengthm1}      & 1,2,3,8,10 & 5 & Magnetic dipole strength \\
\texttt{breakupmodel}    & 1--2     & 2  & Nucleon breakup model \Bstrut\\
\hline
\end{tabular}
\end{table}

In addition to these discrete adjustments, TALYS offers a number of inputs that take continuous parameter adjustments. 
These function as constants that globally adjust model calculations. 
An example of this is the $M^2$ parameters, \texttt{m2constant}, \texttt{m2limit}, and \texttt{m2shift}, which are variables in the calculation of the square matrix element in pre-equilibrium exciton model.
These inputs have a default value of 1 as specified in the TALYS manual, but can be modified to any value (0.1, 100).
Again, given the breadth of covering this full parameter space, adjusted inputs were down-selected to the values provided in \autoref{tab:contmod}.

The user can see from these two examples how adjusting both discrete models and continuous parameters can have significant effects on predicted nuclear interactions and overall residual cross sections.

\begin{longtable}{cccc}
\label{tab:contmod}\\
\caption{Continuous parameter adjustments.}  \\        
\hline\hline
Group & TALYS Keyword & Parameter & Range \\
\hline

\endfirsthead
\caption[]{(continued).}\\
\hline\hline
Group & TALYS Keyword & Parameter & Range\\
\hline
\endhead
\hline \multicolumn{4}{r}{(Continued on next page)}\\
\endfoot

\hline\hline
\endlastfoot

Level Density & \texttt{rspincut} & $\sigma$ &(0.1, 10)\Tstrut\\
Optical & \texttt{avadjust n} & $a_V^n$ &(0.1, 10)\\
& \texttt{avadjust p}   & $a_V^p$    &(0.1, 10) \\
& \texttt{avsoadjust n} & $a_{V,so}^n$&(0.1, 10) \\
& \texttt{avsoadjust p} & $a_{V,so}^p$&(0.1, 10) \\
& \texttt{awadjust n}   & $a_W^n$    &(0.1, 10) \\
& \texttt{awadjust p}   & $a_W^p$    &(0.1, 10) \\
& \texttt{awsoadjust n} & $a_{W,so}^n$&(0.1, 10) \\
& \texttt{awsoadjust p} & $a_{W,so}^p$&(0.1, 10) \\
& \texttt{d1adjust n}   & $d_1^n$    &(0.1, 10) \\
& \texttt{d1adjust p}   & $d_1^p$    &(0.1, 10) \\
& \texttt{d2adjust n}   & $d_2^n$    &(0.1, 10) \\
& \texttt{d2adjust p}   & $d_2^p$    &(0.1, 10) \\
& \texttt{d3adjust n}   & $d_3^n$    &(0.1, 10) \\
& \texttt{d3adjust p}   & $d_3^p$    &(0.1, 10) \\
& \texttt{rcadjust n}   & $r_c^n$    &(0.1, 10) \\
& \texttt{rcadjust p}   & $r_c^p$    &(0.1, 10) \\
& \texttt{rvadjust n}   & $r_V^n$    &(0.1, 10) \\
& \texttt{rvadjust p}   & $r_V^p$    &(0.1, 10) \\
& \texttt{rvsoadjust n} & $r_{V,so}^n$&(0.1, 10) \\
& \texttt{rvsoadjust p} & $r_{V,so}^p$&(0.1, 10) \\
& \texttt{rwadjust n}   & $r_W^n$    &(0.1, 10) \\
& \texttt{rwadjust p}   & $r_W^p$    &(0.1, 10) \\
& \texttt{rwsoadjust n} & $r_{W,so}^n$&(0.1, 10) \\
& \texttt{rwsoadjust p} & $r_{W,so}^p$&(0.1, 10) \\
& \texttt{v1adjust n}   & $v_1^n$    &(0.1, 10) \\
& \texttt{v1adjust p}   & $v_1^p$    &(0.1, 10) \\
& \texttt{v2adjust n}   & $v_2^n$    &(0.1, 10) \\
& \texttt{v2adjust p}   & $v_2^p$    &(0.1, 10) \\
& \texttt{v3adjust n}   & $v_3^n$    &(0.1, 10) \\
& \texttt{v3adjust p}   & $v_3^p$    &(0.1, 10) \\
& \texttt{v4adjust n}   & $v_4^n$    &(0.1, 10) \\
& \texttt{v4adjust p}   & $v_4^p$    &(0.1, 10) \\
& \texttt{vso1adjust n} & $v_{so1}^n$&(0.1, 10) \\
& \texttt{vso1adjust p} & $v_{so1}^p$&(0.1, 10) \\
& \texttt{vso2adjust n} & $v_{so2}^n$&(0.1, 10) \\
& \texttt{vso2adjust p} & $v_{so2}^p$&(0.1, 10) \\
& \texttt{w1adjust n}   & $w_1^n$    &(0.1, 10) \\
& \texttt{w1adjust p}   & $w_1^p$    &(0.1, 10) \\
& \texttt{w2adjust n}   & $w_2^n$    &(0.1, 10) \\
& \texttt{w2adjust p}   & $w_2^p$    &(0.1, 10) \\
& \texttt{w3adjust n}   & $w_3^n$    &(0.1, 10) \\
& \texttt{w3adjust p}   & $w_3^p$    &(0.1, 10) \\
& \texttt{w4adjust n}   & $w_4^n$    &(0.1, 10) \\
& \texttt{w4adjust p}   & $w_4^p$    &(0.1, 10) \\
& \texttt{wso1adjust n} & $w_{so1}^n$&(0.1, 10) \\
& \texttt{wso1adjust p} & $w_{so1}^p$&(0.1, 10) \\
& \texttt{wso2adjust n} & $w_{so2}^n$&(0.1, 10) \\
& \texttt{wso2adjust p} & $w_{so2}^p$&(0.1, 10) \\
Pre-Equilibrium & \texttt{cbreak a} & $c_{\text{break}}^a$ &(0.1, 10) \\
& \texttt{cbreak d} & $c_{\text{break}}^d$ &(0.1, 10) \\
& \texttt{cbreak p} & $c_{\text{break}}^p$ &(0.1, 10) \\
& \texttt{cbreak t} & $c_{\text{break}}^t$ &(0.1, 10) \\
& \texttt{cknock a} & $c_{\text{knock}}^a$ &(0.1, 10) \\
& \texttt{cknock d} & $c_{\text{knock}}^d$ &(0.1, 10) \\
& \texttt{cknock p} & $c_{\text{knock}}^p$ &(0.1, 10) \\
& \texttt{cknock t} & $c_{\text{knock}}^t$ &(0.1, 10) \\
& \texttt{cstrip a} & $c_{\text{strip}}^a$ &(0.1, 10) \\
& \texttt{cstrip d} & $c_{\text{strip}}^d$ &(0.1, 10) \\
& \texttt{cstrip p} & $c_{\text{strip}}^p$ &(0.1, 10) \\
& \texttt{cstrip t} & $c_{\text{strip}}^t$ &(0.1, 10) \\
 & \texttt{m2constant} & $C_1$ & (0.1,100) \\
 & \texttt{m2limit} & $C_2$ & (0.1,100) \\
 & \texttt{m2shift} & $C_3$ & (0.1,100) \\
 & \texttt{rgamma} & $g$ & (0.1,100)\\
 & \texttt{rnunu} & $r_{\nu\nu}$ & (0.1,100)\\
 & \texttt{rnupi}& $r_{\nu\pi}$ & (0.1,100)\\
 & \texttt{rpinu} & $r_{\pi\nu}$ & (0.1,100)\\
 & \texttt{rpipi} & $r_{\pi\pi}$ & (0.1,100)\Bstrut\\

\end{longtable}

For many of these inputs, both discrete and continuous, TALYS will also allow for fine-tuning based on the outgoing residual product.
However, given the scope of these calculations, this functionality was not explored.

Previous efforts focused on tuning parameters to achieve a best fit for a given experiment or specific residual product cross sections \cite{fox_measurement_2021,morrell_measurement_2026, Apgar2026SnTe}.
A best fit for a given isotope and residual product may be achievable with significant computational power, but it does not tell the full story for all residual products from that reaction, nor does it necessarily apply to another isotope and its residual products.
An exhaustive search of all combinations of all models and parameters is not physically possible due to computational resources. 
However, an informed evaluation of the available models and parameters, computational bandwidth, and existing literature \cite{alhassan_iterative_2022} allowed for this down-selection of parameters for adjustment. 
This introduces an additional bias, but note that even the selection of TALYS 2.0 over other versions, or other theoretical modeling codes is in itself a statistical bias.
A broader study of these biases is being explored by Alhassan \etal \cite{alhassan_bayesian_2024}.

\subsection{Prior Sampling and Bayesian Updating}

\begin{table*}

\caption{Notation used in this section.}
\label{tab:notation}
\begin{tabular}{ccc}
\hline \hline
\textbf{Use} & \textbf{Symbol} & \textbf{Description} \\

\hline
Indices and subscripts & $i$ & Parameter index ($1$ to $N_{\mathrm{params}}$)\\
& $j$ & Sample/model index ($1$ to $N_{\mathrm{samples}}$) \\
& $c$ & Reaction channel index ($1$ to $N_c$) \\
& $d$ & Dataset index within a channel ($1$ to $N_d$) \\
& $pt$ & Data point index within a dataset ($1$ to $N_{pt}$) \\

Parameters and distributions & $\theta$, $\theta_i$ & Parameter vector and individual parameter \\
& $p(\theta)$ & Prior density  \\
& $q(\theta)$ & Proposal density \\
& $\mu$, $\sigma$ & Prior log-space mean and standard deviation of initial generation \\
& $\mu_{\mathrm{post}}, \sigma_{\mathrm{post}}$ & Posterior log-space mean and standard deviation of subsequent generations \\
& $[\mathrm{lo}, \mathrm{hi}]$ & Prior truncation bounds\\
& $[g_{\mathrm{lo}}, g_{\mathrm{hi}}]$ & Proposal truncation bounds derived from posterior data \\

Chi-square notation & $\chi^2_j$ & Chi-square value for sample/model $j$ \\
 & $\chi^2_{\mathrm{T}}$ & Chi-square for established metric \\
& $\chi^2_{\tau,j}$ & Normalized chi-square ratio \\

Cross-sections & $\sigma^{(c,d,pt)}_{\mathrm{model},j}$ & Predicted cross-section for model $j$ at point $(c,d,pt)$ \\
& $\sigma^{(c,d,pt)}_{\mathrm{exp}}$ & Experimental cross-section from EXFOR \\
& $\delta \sigma^{(c,d,pt)}_{\mathrm{exp}}$ & Experimental uncertainty (minimum 15\% enforced) \\

Weights & $\mathcal{L}_j$ & Likelihood for model $j$ \\
& $w_j$ & Normalized posterior weight for model $j$ \\
\hline \hline
\end{tabular}
\end{table*}

\subsubsection{Generation of a Prior Distribution}

An initial pass of 3000 TALYS calculations was generated to define the prior distribution over both discrete models and continuous parameters, allowing for a broad exploration of the physics-based models and tunable model constants. 
Discrete models were chosen at random, while continuous parameters were sampled using Sobol sequences with scipy.stats.qmc.Sobol \cite{Virtanen2020SciPy, JoeKuo2008Sobol} to improve coverage over the high-dimensional parameter space, reducing clustering in comparison to simple random draws. Scrambling was enabled to add randomness while preserving the low-discrepancy properties\cite{owen_scrambling_1998}. These were drawn from a truncated lognormal distribution centered on the default TALYS parameter values for each parameter and sampled in log space with parameters $\mu$=0 and $\sigma$=0.5. Here, $\mu$=0 centers the distribution around the TALYS default parameter value and $\sigma$=0.5 creates a moderate spread around that default. Truncation was enforced to ensure that the parameter values would not fall outside the allowed values within TALYS. This establishes a prior distribution as follows in \autoref{eq:priortheta},\ref{eq:prior} for continuous parameters and \autoref{eq:priordisc} for discrete parameters, with the overall prior distribution being a product of these.
\begin{equation}
    p(\theta) = p(\theta_{\text{cont}}) \cdot p(\theta_{\text{disc}}) = \prod_{i=1}^{N_{\text{cont}}} p_i(\theta_i) \cdot \prod_{k=1}^{N_{\text{disc}}} p_k(\theta_k)
    \label{eq:priortheta}
\end{equation}
\begin{equation}
p_i(x) = \frac{1}{x \sigma \sqrt{2\pi}} \cdot \frac{\exp\left( -\frac{(\ln x - \mu)^2}{2\sigma^2} \right)}{\Phi\left( \frac{\ln(\text{hi}_i) - \mu}{\sigma} \right) - \Phi\left( \frac{\ln(\text{lo}_i) - \mu}{\sigma} \right)}
    \label{eq:prior}
\end{equation}
\begin{equation}
p_k(\theta_k) = \frac{1}{|\mathcal{S}_k|}
\label{eq:priordisc}
\end{equation}
In \autoref{eq:priordisc}, $|\mathcal{S}_k|$ is the set of potential values for the discrete parameter $k$, as provided in \autoref{tab:discrete_params}.
In contrast to a uniform sampling, which would give equal probability to all potential parameter values, a truncated lognormal distribution would limit the number of nonphysical parameter combinations. The probability density would concentrate sampling around the default values while still allowing some exploration of the parameter value regions far from the defaults.

\subsubsection{Establishing a Goodness-of-Fit Metric}

The resulting outputs of this initial set of TALYS calculations were evaluated by comparing the predicted residual product cross sections against experimental measurements, producing a reduced $\chi^2$ statistic.
Because the aggregation of discrepancies across channels, datasets, and data points can influence the overall assessment of a given model's performance, multiple definitions of $\chi^2$ were investigated. This work explored four principal forms of this metric and four methods of normalization on these forms. In the following equations, $N_c$ is the number of experimental channels, $N_d$ is the number of datasets for that channel, $N_pt$ is the number of data points from the dataset, $\sigma_{model}^{(c,d,pt)}$ is the predicted cross section for a given TALYS calculation, $\sigma_{exp}^{(c,d,pt)}$ is the experimental cross section measurement, and $\delta\sigma_{exp}^{(c,d,pt)}$ is the experimental measurement uncertainty.
For experimental measurements with no uncertainty, or an uncertainty less than 5\%, an assumed 15\,\% uncertainty was enforced.
In the first approach, the reduced $\chi^2$ was calculated in each dataset, then aggregated within each reaction channel, and finally averaged across all channels, as seen in \autoref{eq:nested}
\begin{equation}
\begin{aligned}
& \chi^2_{\text{nested}} = \frac{1}{N_c} \sum_{c=1}^{N_c} \\ & \left[ \frac{1}{N_d^{(c)}} \sum_{d=1}^{N_d^{(c)}} \left( \frac{1}{N_{pt}^{(c,d)}} \sum_{pt=1}^{N_{pt}^{(c,d)}} \frac{(\sigma^{(c,d,pt)}_{\text{model}} - \sigma^{(c,d,pt)}_{\text{exp}})^2}{(\delta\sigma^{(c,d,pt)}_{\text{exp}})^2} \right) \right]
    \label{eq:nested}
    \end{aligned}
\end{equation}

The second approach calculated across each dataset, aggregated for each point irrespective of channel, and averaged across all datasets, as seen in \autoref{eq:dataset_balanced}
\begin{equation}
\chi^2_{\text{dataset}} = \frac{1}{N_d^{\text{total}}} \sum_{d=1}^{N_d^{\text{total}}} \left( \frac{1}{N_{pt}^{(d)}} \sum_{pt=1}^{N_{pt}^{(d)}} \frac{(\sigma^{(d,pt)}_{\text{model}} - \sigma^{(d,pt)}_{\text{exp}})^2}{(\delta\sigma^{(d,pt)}_{\text{exp}})^2} \right)
    \label{eq:dataset_balanced}
\end{equation}

In the third approach, each point in a channel, irrespective of datasets, were aggregated and averaged across each channel, seen in \autoref{eq:channel_balanced}
\begin{equation}
\chi^2_{\text{channel-pt}} = \frac{1}{N_c} \sum_{c=1}^{N_c} \left( \frac{1}{N_{pt}^{(c)}} \sum_{pt=1}^{N_{pt}^{(c)}} \frac{(\sigma^{(c,pt)}_{\text{model}} - \sigma^{(c,pt)}_{\text{exp}})^2}{(\delta\sigma^{(c,pt)}_{\text{exp}})^2} \right)
    \label{eq:channel_balanced}
\end{equation}

Finally, a $\chi^2$ metric was calculated in which all points, irrespective of dataset or channel, was aggregated and averaged - here seen in \autoref{eq:pointwise}.
\begin{equation}
\chi^2_{\text{pointwise}} = \frac{1}{N_{pt}^{\text{total}}} \sum_{pt=1}^{N_{pt}^{\text{total}}} \frac{(\sigma^{(pt)}_{\text{model}} - \sigma^{(pt)}_{\text{exp}})^2}{(\delta\sigma^{(pt)}_{\text{exp}})^2}
    \label{eq:pointwise}
\end{equation}

\begin{figure*}
    \centering
    \includegraphics[width=0.9\linewidth]{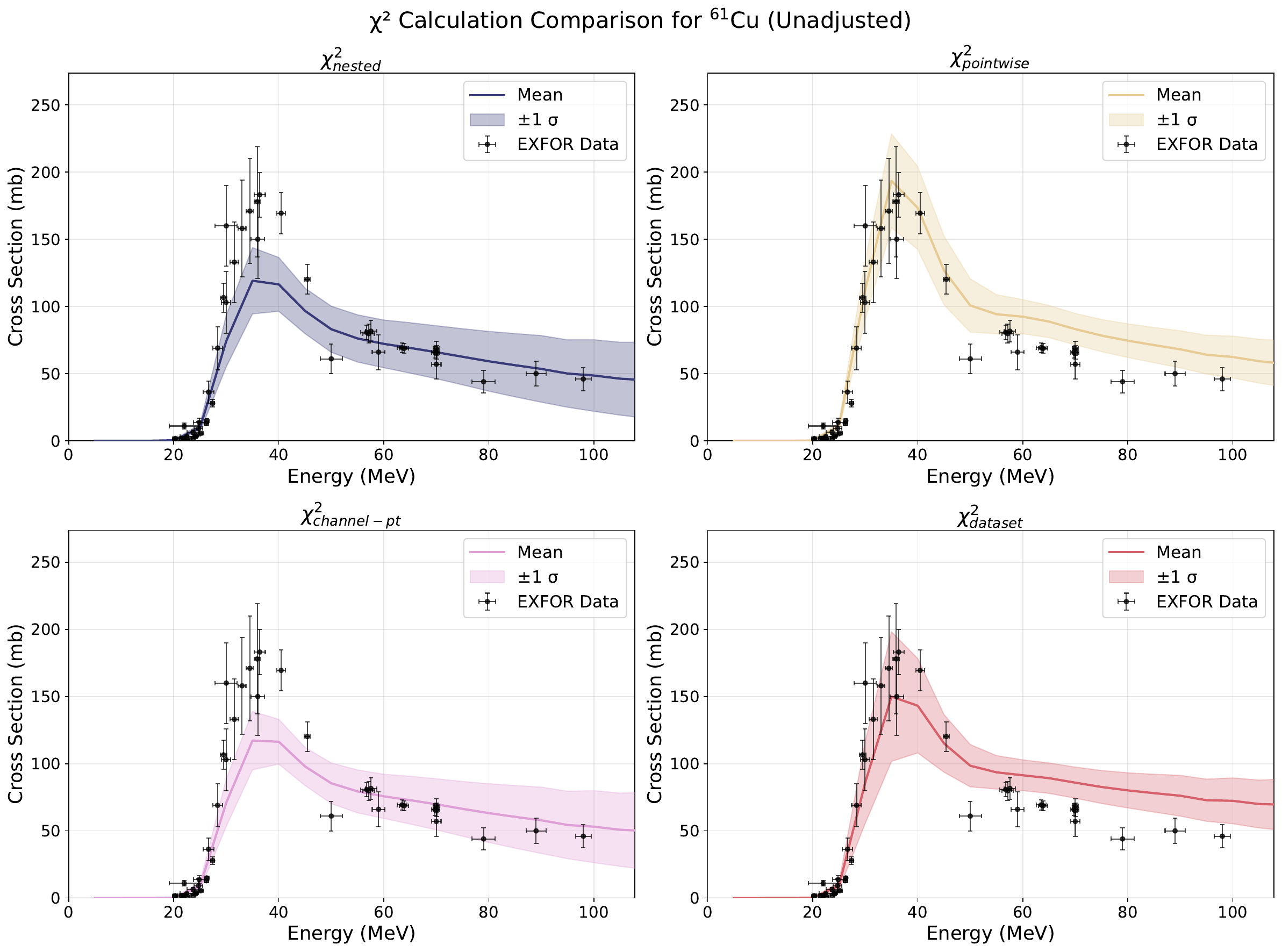}
    \caption{Visualization of competing $\chi^2$ metrics, along with the overall best $\chi^2$ value across all inputs, the ESS, and the percent of experimental datapoints contained across all channels. These figures specifically illustrate the residual product \ce{^{61}Cu} for visualization.}
    \label{fig:61Cuchi2unadjusted}
\end{figure*}

A visualization of these metrics can be seen in \autoref{fig:61Cuchi2unadjusted}, showing the coverage for each method on a sample product, \ce{^{61}Cu}. 

Of note is that the ESS for all of these methods is incredibly small, indicating that very few TALYS calculations actually contribute to determining this established mean and standard deviation. 
To provide context to these values and increase ESS, so that more inputs contribute to the calculations, a normalization to established metrics was considered, including the performance of TENDL-2023, the default TALYS 2.0 calculation, and the default TALYS 2.0 calculation with the maximum excitation energy bin setting of bins=100.
For a given result, we calculate the ratio of the reduced $\chi^2$ to that of the reference calculation. In this way, a normalized value below 1.0 indicates an improvement relative to the reference, whereas a value above 1.0 indicates poorer performance. 

For example, \autoref{eq:nested} could be modified such that establish a goodness of fit for each theoretical model for experimental channels using $\chi^2_{\tau(j,k),c}$, here in \autoref{eq:aggdp}.
\begin{equation}
    \chi^2_{j,c} = \frac{1}{N_d^{(c)}} \sum_{d=1}^{N_d^{(c)}} \left( \frac{1}{N_{pt}^{(c,d)}} \sum_{pt=1}^{N_{pt}^{(c,d)}} \frac{(\sigma^{(c,d,pt)}_{\text{model},j} - \sigma^{(c,d,pt)}_{\text{exp}})^2}{(\delta\sigma^{(c,d,pt)}_{\text{exp}})^2} \right)
\label{eq:aggdp}
\end{equation}

We then normalize to $\chi^2$ values for each channel based on an established metric $\chi^2_{T,c} $ to arrive at a $\chi^2$ ratio for each model, $\chi^2_{\tau,j}$, calculated in \autoref{eq:normalizedchi2}
\begin{equation}
    \chi^2_{\tau,j} = \frac{1}{N_c} \sum_{c=1}^{N_c} \frac{\chi^2_{j,c}}{\chi^2_{T,c}}
    \label{eq:normalizedchi2}
\end{equation}

In the case where channels are not individually aggregated, the total $\chi^2$ value of the established metrics was used as the denominator for the ratio.

The different calculation metrics were also compared in their effective sample size (ESS), which provides guidance on the percent of samples contributing significantly to the posterior distribution calculation (See \autoref{eq:ess}) as well as the percent coverage of EXFOR experimental datapoints within the posterior mean $\pm 1\sigma$. These results are provided in \autoref{tab:ess_containment_gen0_expanded}.

These $\chi^2$ ratios provide an interpretable measure of performance relative to the reference calculations. 
The result of these comparisons can be seen in \autoref{tab:chi2metrics}.
\begin{table}
\caption{$\chi^2$ best fits across each calculation metric.}
\label{tab:chi2metrics}
\begin{tabular}{lcccc}
\hline \hline
Metric &   Unadjusted  & TENDL & TALYS & bins=100 \\
\hline
$\chi^2_{nested}$ & 33.026 & 0.274 & 0.679 & 0.823 \\
$\chi^2_{dataset}$ & 32.896 & 1.226 & 2.623 & 1.265 \\
$\chi^2_{channel-pt}$ & 38.785 & 1.275 & 2.430 & 1.343 \\
$\chi^2_{pointwise}$ & 29.838 & 0.976 & 2.196 & 1.042 \\
\hline \hline
\end{tabular}
\end{table}

\begin{table*}[t]
\centering
\caption{ESS and containment for Generation 0. Coverage is the percentage of EXFOR points within posterior $\pm 1\sigma$.}
\label{tab:ess_containment_gen0_expanded}
\renewcommand{\arraystretch}{1.1}
\begin{tabular}{lcc cc cc cc}
\hline \hline
& \multicolumn{2}{c}{Unadjusted} 
& \multicolumn{2}{c}{TENDL 2023}
& \multicolumn{2}{c}{TALYS 2.0} 
& \multicolumn{2}{c}{bins=100} \\
\cmidrule(lr){2-3} \cmidrule(lr){4-5} \cmidrule(lr){6-7} \cmidrule(lr){8-9}
Metric & ESS\,\% & Coverage\,\% & ESS\,\%& Coverage\,\% & ESS\,\% & Coverage\,\% & ESS\,\% & Coverage\,\% \\
\midrule
$\chi^2_{nested}$            & 0.1 & 57.2 & 49.1 & 71.3 & 56.1 & 74.9 & 46.3 & 70.9 \\
$\chi^2_{dataset}$  & 0.2 & 69.7 & 20.5 & 70.4 & 36.0 & 71.8 & 35.4 & 71.6 \\
$\chi^2_{channel-pt}$  & 0.2 & 58.0 & 30.0 & 67.1 & 41.3 & 69.1 & 40.5 & 69.0 \\
$\chi^2_{pointwise}$       & 0.2 & 74.9 & 16.8 & 77.5 & 37.7 & 75.8 & 35.9 & 75.8 \\
\hline \hline
\end{tabular}
\end{table*}

\begin{figure*}
    \centering
    \includegraphics[width=.8\linewidth]{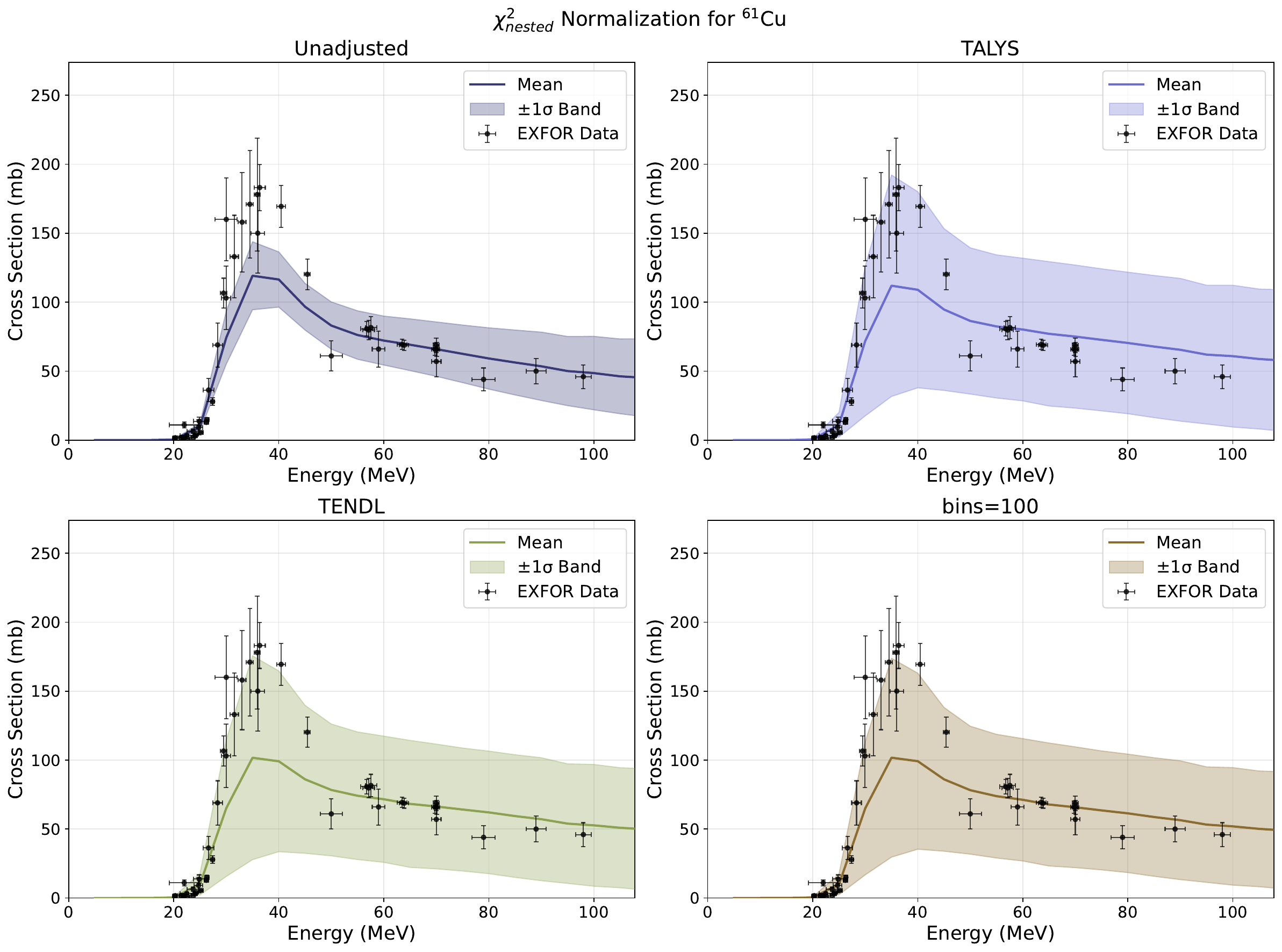}
    \caption{Unadjusted and normalized $\chi^2_{nested}$ results, showing the improvement in ESS through normalization and increased containment, at the loss of finer uncertainty bands, for the residual product \ce{^{61}Cu}.}
    \label{fig:nestednormalization}
\end{figure*}

The $\chi^2_{nested}$ calculation was settled on for this effort. There are both benefits and drawbacks to this assignment of a $\chi^2$ value, in that every channel, every experiment, and every data point is equally considered, so a channel with hundreds of data points among many datasets would be equally considered against a channel with only a few datapoints and datasets. 
The purpose of this is to avoid over-weighting any specific channel, as we seek to find the best holistic fit for all measured residual products, rather than a fit for specific datasets and channels alone. A second visualization of \ce{^{61}Cu} using these methods shown in \autoref{fig:nestednormalization}.

The next decision focused on whether to use the unadjusted values, or to normalize them to the existing evaluated standards. Again, a comparison of these options, each included benefits and drawbacks.
The unadjusted dataset was declined, as it resulted in heavily over-weighting a few samples.
Using the default TALYS 2.0 calculation as a ratio was declined as well, since the calculations performed in this work by default used \texttt{bins}=100. TENDL-2023 was declined as well, as it could introduce the biases of previously evaluated nuclear data.
Introducing these biases contradicts the purpose of this analysis, as we are not searching for a single best fit of model and parameter combinations, but rather for establishing a global uncertainty band.

In summary, and going forward, we will refer to $\chi^2_{nested}$, normalized to TALYS 2.0 with \texttt{bins}=100, simply as $\chi^2$ for simplicity, as it was chosen as the metric for evaluation.
This choice eliminates the bias on large datasets or specific channels that have been well-characterized as well as preventing overfitting on an incredibly small fraction of TALYS runs. We also seek to compare equally to our inputs, using the highest number of excitation energy bins against the default TALYS results, and not to replicate or be impacted by previously evaluated nuclear data libraries.
\\

This work does not intend to serve as a nuclear data evaluation, but rather as an evaluation of reaction modeling.
If an evaluation was focused on only a few residual products, one could weigh these more heavily. Similarly, a dataset with hundreds of evaluated datapoints could be considered more established when compared to one with just a few datapoints from a single experiment.

The goal is not to find the best fit for these well characterized channels, nor is it to perform an artificial evaluation of EXFOR datasets, but to determine the uncertainty bands for reaction channels with little, or even no measured data but establishing uncertainties on TALYS parameters. 

In this case, aside from enforcing additional uncertainties to experimental measurements, we consider all measured independent cross sections to be of equal value, agnostic of the experimental data, residual data, or number of data points. 

\subsubsection{Likelihood-Based Model Weighting}
With our overall $\chi^2_{\tau(j,k)}$ for each theoretical model set $\vec{M_j}$ and parameter values $\vec{p_k}$, we can establish, assuming normally distributed experimental errors, the likelihood of the data given the model proportional to 
\autoref{eq:likelihood}.

\begin{equation}
    \mathcal{L}_j = \exp\left( -\frac{\chi^2_{\tau,j}}{2} \right)
    \label{eq:likelihood}
\end{equation}

To convert these likelihoods into normalized sample weights $w_{j,k}$ we define \autoref{eq:weights}

\begin{subequations}
\label{eq:weights}
\begin{align}
    &w_j^{(0)} = \frac{\mathcal{L}_j}{\sum_{a=1}^{N} \mathcal{L}_a} = \frac{\exp(-\chi^2_{\tau,j}/2)}{\sum_{a=1}^{N} \exp(-\chi^2_{\tau,a}/2)}\label{eq:weights0}\\
    & w_j^{(n+1)} = \frac{\mathcal{L}_j \cdot \frac{p(\theta_j)}{q(\theta_j)}}{\sum_{a=1}^{N} \mathcal{L}_a \cdot \frac{p(\theta_a)}{q(\theta_a)}}\label{eq:weights1}
\end{align}
\end{subequations}

where N is the number of TALYS iterations. This transformation, similar to a softmax, results in models with better performance receiving larger weights while still retaining nonzero contribution unless fit is very poor. Rather than a hard rejection threshold, this weighting avoids eliminating potentially informative regions of the parameter space, resulting in a smoother posterior distribution.

The weighting in generations following the initial generation can no longer be considered uniform, so we introduce a new ratio such that $w_i \not\propto L_i\cdot p(\theta_i)$, but rather $w_i\propto L_i\cdot\dfrac{p(\theta_i)}{q(\theta_i)}$, where $p(\theta_i)$ is the initial generation's prior density, defined above in Equations \ref{eq:priortheta} and \ref{eq:prior}, and $q(\theta_i)$ is the proposal density that the sample was drawn from, defined in Equations \ref{eq:proptheta} and \ref{eq:prop}.

\begin{equation}
    q(\theta) = \prod_{i=1}^{N} q_i(\theta_i)
    \label{eq:proptheta}
\end{equation}
\begin{equation}
\begin{aligned}
& q_i(x) = \frac{1}{x \sigma_{\text{post},i} \sqrt{2\pi}} \cdot \\ 
&\frac{\exp\left( -\frac{(\ln x - \mu_{\text{post},i})^2}{2\sigma_{\text{post},i}^2} \right)}{\Phi\left( \frac{\ln(g_{\text{hi},i}) - \mu_{\text{post},i}}{\sigma_{\text{post},i}} \right) - \Phi\left( \frac{\ln(g_{\text{lo},i}) - \mu_{\text{post},i}}{\sigma_{\text{post},i}} \right)}
\label{eq:prop}
\end{aligned}
\end{equation}

\subsubsection{Posterior updating and iterative generation of new samples}
For discrete model parameters, interpolation is not meaningful, so the value of these were fixed after the initial generation based on the best-performing sample, defined by the sample with greatest posterior weight. 

Continuous parameters for subsequent generations were no longer sampled blindly, but rather on the performance of the prior based on an updated distribution from the weighted model statistics.

The weighted posterior mean of a parameter $x_i$ is defined by \autoref{eq:parammean}.
\begin{equation}
    \label{eq:parammean}
    \mu_i = \frac{\sum_{j=1}^{N} w_j \cdot x_{ij}}{\sum_{j=1}^{N} w_j}
\end{equation}

with a corresponding weighted variance calculated using \autoref{eq:paramvar}.
\begin{equation}
\label{eq:paramvar}
\sigma_i^2 = \frac{\sum_{j=1}^{N} w_j \cdot (x_{ij} - \mu_i)^2}{\sum_{j=1}^{N} w_j}
\end{equation}

For later generations, new samples were drawn from distributions centered on the weighted mean and spread determined by the weighted standard deviation. 
This method progressively concentrates values in regions of higher likelihood while preserving variance for continued parameter value exploration.
Again, since these parameters function as multipliers, they were sampled using truncated lognormal distributions and Sobol sampling to sample efficiently.

These subsequent generations of parameter sets were passed to TALYS, with the resulting theoretical cross section results again compared with experimental data to arrive at $\chi^2$ values and posterior weights. 

\subsubsection{Criteria for Convergence}
This cycle of resampling was repeated iteratively, with successive generations fine-tuning the mean and uncertainty associated with the TALYS parameter set. 
This is done to reduce the uncertainty predicted in NDS-2022 and, not arriving at any specific set of parameter values for an optimized TALYS calculation, but instead to establish an uncertainty band in the predicted cross sections and parameters.
Although each individual run will experience some measure of variation, convergence can be improved through large sample sizes and multiple generations. 
With this approach, it was determined that generations of 1000 TALYS calculations sufficiently covered the exploration space.

Convergence was based on a number of criteria.
The first criterion was a decreasing return on investment per generation based on the best $\chi^2$ value. Convergence could be considered when improvement over the current and prior generation was $<5\%$.
Another criterion was based on the median $\chi^2$ value, when the reduction was $<10\%$.
The effective sample size, or ESS, defined in \autoref{eq:ess} provided guidance as well. This metric is useful to determine the number of samples that are actually having a significant difference on calculating the posterior distribution. The metric used here required at least 10\% of samples to be contributing, as any less would be computationally expensive for very little gain.
\begin{equation}
\text{ESS} = \frac{1}{\sum_{j=1}^{N} (w_j)^2}
\label{eq:ess}
\end{equation}
Finally, the percentage of experimental datapoints that were covered in the $\pm 1-\sigma$ uncertainty band was required to stay at or above roughly 68\%, including added uncertainties for experimental datapoints with $<5\%$ or no uncertainty listed. This metric was used to ensure that data was not overfitted.

\subsubsection{Computational Resources}

This research used the Lawrencium computational cluster resource provided by the IT Division at the Lawrence Berkeley National Laboratory (Supported by the Director, Office of Science, Office of Basic Energy Sciences, of the U.S. Department of Energy under Contract No. DE-AC02-05CH11231). TALYS computations leveraged the lr8 partition on Lawrencium, with each node equipped with an AMD EPYC 9534 CPU and 128 cores, and 768 GB memory. Data was stored in PostgreSQL version 16.3 databases. Local analysis was performed on a high-performance workstation equipped with an Apple M3 Max system-on-a-chip (SoC) with 64 GB RAM, with Python 3.13 for analysis. Total evaluations ultimately utilized 8000 TALYS runs for analysis. Data analysis and code development were assisted by Claude Code \cite{claude2026} and ChatGPT \cite{openai_chatgpt_2026}, with all generated code and analysis verified by the authors.







    \section{\label{sec:Results}Results and discussion\protect}
        \subsection{Convergence of Data Over Generations}

\subsubsection{Discrete Model Optimization}
After the initial exploratory run of 3000 inputs, the best-performing input was used to freeze discrete model parameters. 
In comparison to TALYS 2.0 default values, the following models were chosen, as seen in \autoref{tab:final_params}.

\begin{table}

\caption{Discrete parameter adjustments after initial TALYS runs.}
\label{tab:final_params}
\centering
\begin{tabular}{l l l l}

\hline \hline
Family & Parameter & Default & Best File \\
\hline
Gamma & \texttt{strength} & 9 & 3 \\
Gamma & \texttt{strengthm1} & 3 & 10 \\
Level Density & \texttt{kvibmodel} & 2 & 1 \\
Level Density & \texttt{ldmodel} & 1 & 3 \\
Level Density & \texttt{parity} & 0 & 1 \\
Level Density & \texttt{spincutmodel} & 1 & 2 \\
Pre-equilibrium & \texttt{breakupmodel} & 1
 & 1 \\
Pre-equilibrium & \texttt{gshell} & 0 & 1 \\
Pre-equilibrium & \texttt{preeqmode} & 2 & 3 \\
Pre-equilibrium & \texttt{preeqspin} & 2 & 3 \\
\hline \hline
\end{tabular}
\end{table}

\subsubsection{Iterations Over Continuous Parameters}
Following the methodology discussed above, a total of 5 additional generations were run after the initial iteration. The summary statistics over these generations can be seen in \autoref{tab:generation_summary}. $\chi^2$ values utilized all existing datasets for independent cross sections for \ce{^{nat}Cu}(p,x) reactions, which at the time of research was 34 datasets, covering 24 independent products comprising over 2600 data points \cite{jung_cross_1992,michel_cross_1997,gruetter_excitation_1982,hermanne_excitation_1999,voyles_excitation_2018, voyles_proton-induced_2021, liu_excitation_2021,orth_pion-induced_1978,garrido_new_2016,buthelezi_excitation_2006,szelecsenyi_excitation_2001,al-saleh_excitation_2006, yashima_measurement_2003, heydegger_thin-target_1972,haller_relative_1961, williams_excitation_1967,cervenak_new_2020,kuhnhenn_thin_2001,griswold_measurement_2013, greenwood_measurement_1984,anwer_excitation_2022,lagarde-simonoff_spallation_1975,shahid_measurement_2015,fox_investigating_2021,uddin_experimental_2004,khandaker_excitation_2007,kopecky_proton_1985,takacs_new_2002,graves_nuclear_2016,mills_experimental_1992,siiskonen_excitation_2009,aleksandrov_excitation_1987,yoshizawa_isotope_1976,titarenko_nuclide_2003}.
While the change in the \textit{best} $\chi^2$ value did not improve significantly, the median $\chi^2$ saw a 71.7\% improvement over the 5 generations.
Effective sample size was maintained fairly consistently, with containment also remaining within acceptable bounds. 
In the interest of completion, a sixth generation was run, which saw the effective sample size fall to only 0.7\%, and was therefore excluded from analysis.

\begin{table}[t]
\caption{Generation-by-generation summary of model performance and importance-sampling diagnostics.}
\label{tab:generation_summary}
\begin{ruledtabular}
\begin{tabular}{cccccc}
Gen. &
Best $\chi^2$ &
Weighted mean $\chi^2$ &
Median $\chi^2$ &
ESS (\%) &
Containment \\
\hline
0 & 0.8234 & N/A   & 4.735 & 46\%   & 71.8\% \\
1 & 0.8284 & 2.054 & 2.202 & 59\%  & 70.8\% \\
2 & 0.8481 & 1.950 & 1.834 & 36\%  & 72.6\% \\
3 & 0.8016 & 1.781 & 1.603 & 28\%  & 69.5\% \\
4 & 0.8357 & 1.730 & 1.448 & 15\%  & 70.1\% \\
5 & 0.8041 & 1.546 & 1.334 & 15\%  & 66.6\% \\
6 & 0.8084 & 1.484 & 1.227 & 0.7\% & ---    \\
\end{tabular}
\end{ruledtabular}
\end{table}

\begin{figure}
    \centering
    \includegraphics[width=1\linewidth]{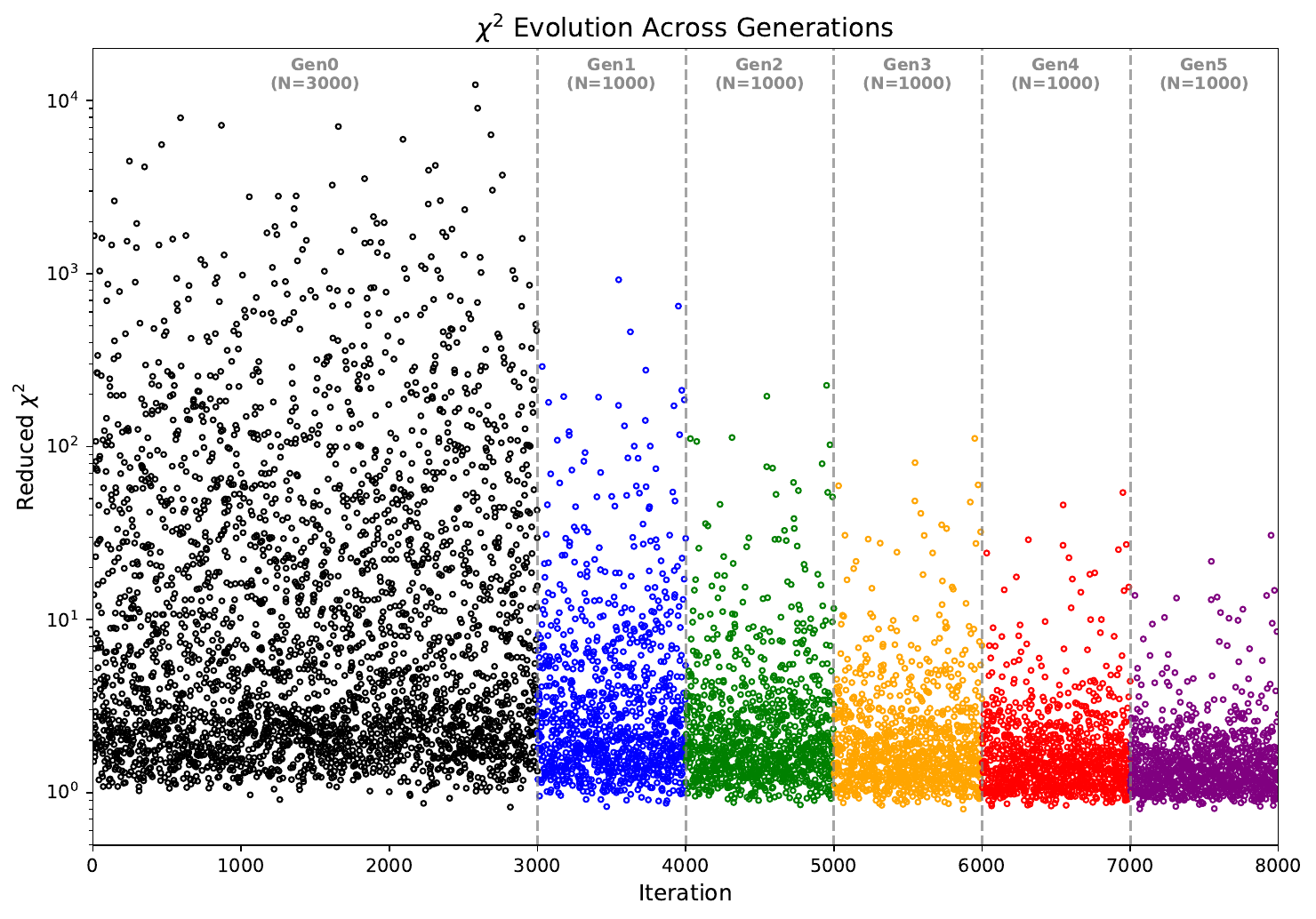}
    \caption{Distribution of $\chi^2$ values across generations.}
    \label{fig:chi2gens}
\end{figure}

This can be visualized in \autoref{fig:chi2gens}, showing a gradual convergence to more accurate $\chi^2$ values over successive generations.

\begin{figure}
    \centering
    \includegraphics[width=1\linewidth]{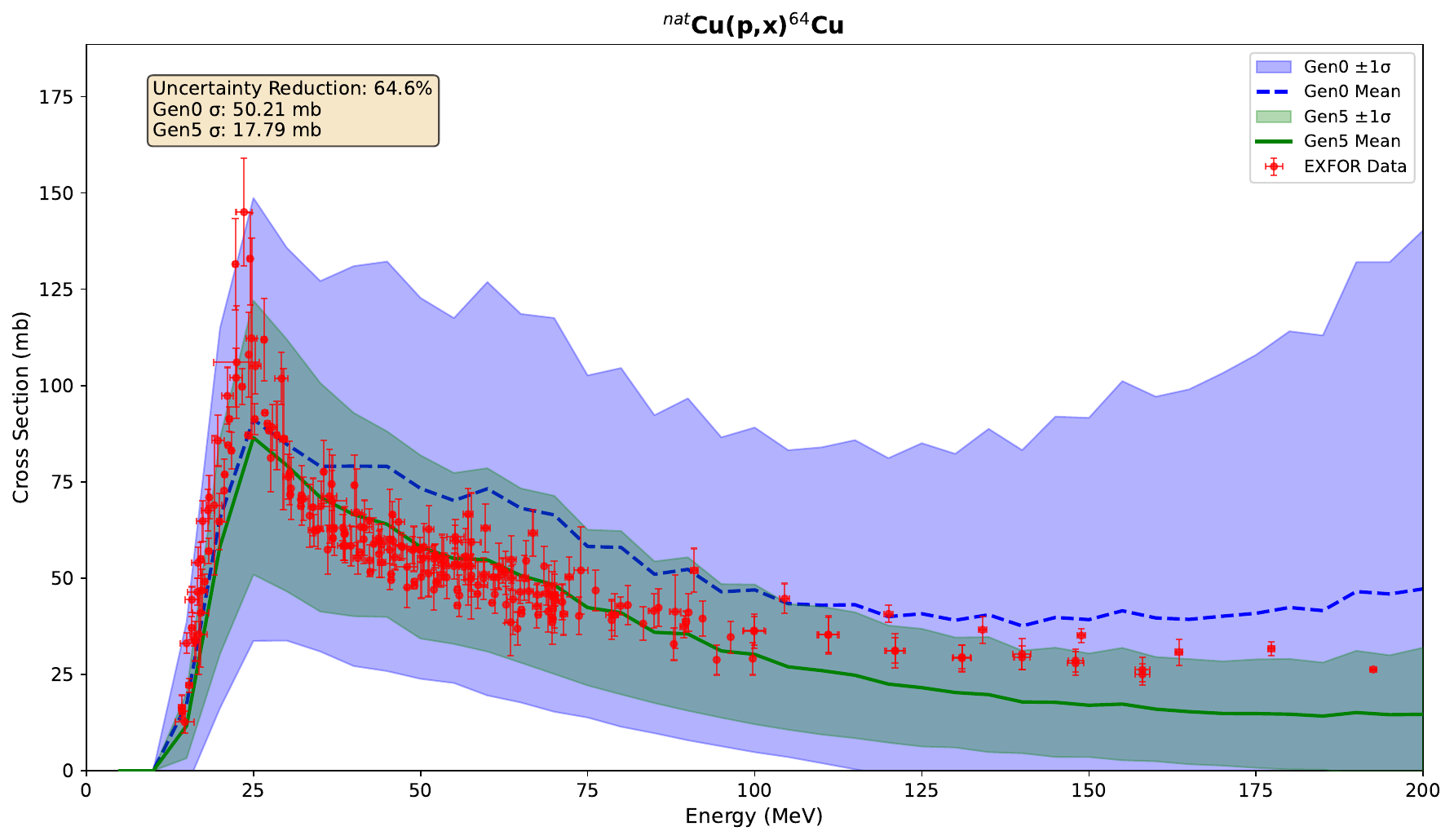}
    \caption{Visualization of the reduction in uncertainty band and improving prediction for the independent residual product, \ce{^{64}Cu}.}
    \label{fig:64Cu}
\end{figure}

At the conclusion of generation 5, residual product cross sections were examined individually to assess the shift from the default and the uncertainty reduction. Visual examples of this are shown in \autoref{fig:64Cu}.

\begin{figure}[htbp]
    \centering
    \includegraphics[width=1\linewidth]{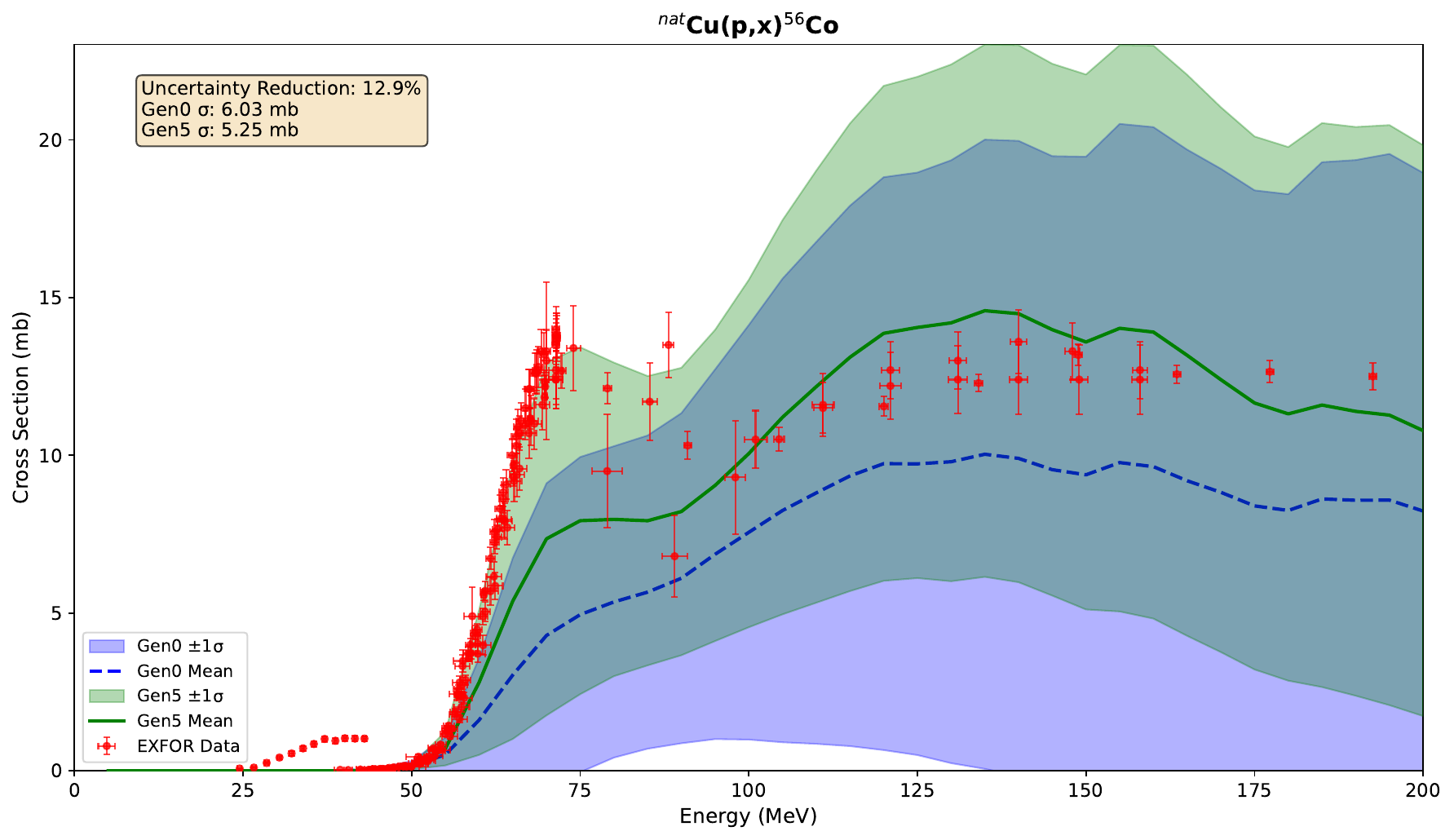}
    \caption{Visualization of the reduction in uncertainty band and improving prediction for the independent residual product, \ce{^{56}Co}.}
    \label{fig:56Co}
\end{figure}

It is important to note that not all residual products demonstrated as great a reduction in uncertainty, as seen in \autoref{fig:56Co} - in fact, several experienced an increase in uncertainty. A breakdown by isotope is available in \autoref{tab:independent_validation}.

\setlength{\LTcapwidth}{\columnwidth}
\begin{longtable}{lcccc}

\caption{Independent cross section validation results comparing Gen. 0 (initial posterior) and Gen. 5 (refined posterior). Columns show isotope, number of EXFOR data points, uncertainty reduction from Gen. 0 to Gen. 5, and containment (Cnt.), the percent of EXFOR points within $\pm 1\sigma$ posterior bands.}
\label{tab:independent_validation}
\\
\hline\hline
Isotope & N$_{\rm pts}$ & $\sigma_{\rm Gen. 5}/\sigma_{\rm Gen. 0}$ & Cnt.$_{\rm Gen. 0}$  & Cont.$_{\rm Gen. 5}$  \\
\hline
\endfirsthead
\caption[]{(continued).}\\
\hline\hline
Isotope & N$_{\rm pts}$ & $\sigma_{\rm Gen. 5}/\sigma_{\rm Gen. 0}$ & Cnt.$_{\rm Gen. 0}$ & Cont.$_{\rm Gen. 5}$  \\
\hline
\endhead
\hline \multicolumn{5}{r}{(Continued on next page)}\\
\endfoot

\hline\hline
\endlastfoot

\ce{^{65}Zn_{(i)}}     & 649 & 0.614 & 97.2  & 90.8 \Tstrut\\
\ce{^{62}Zn_{(i)}}     & 431 & 0.737 & 98.8  & 94.7 \\
\ce{^{57}Co_{(i)}}     & 274 & 0.730 & 14.2  & 12.0 \\
\ce{^{54}Mn_{(i)}}     & 269 & 0.615 & 11.9  & 1.1 \\
\ce{^{64}Cu_{(i)}}     & 228 & 0.354 & 100.0 & 93.9 \\
\ce{^{63}Zn_{(i)}}     & 222 & 0.668 & 95.5  & 95.9 \\
\ce{^{56}Co_{(i)}}     & 183 & 0.871 & 43.2  & 74.3 \\
\ce{^{58}Co_{(i)}}     & 107 & 0.530 & 31.8  & 2.8 \\
\ce{^{60}Co_{(i)}}     & 79  & 0.433 & 74.7  & 30.4 \\
\ce{^{61}Cu_{(i)}}     & 45  & 0.498 & 82.2  & 73.3 \\
\ce{^{57}Ni_{(i)}}     & 14  & 1.310 & 78.6  & 78.6 \\
\ce{^{61}Co_{(i)}}     & 13  & 0.293 & 61.5  & 0.0 \\
\ce{^{46}Sc_{(i)}}     & 12  & 0.684 & 50.0  & 41.7 \\
\ce{^{58m}Co_{(i)}} & 12  & 0.570 & 33.3  & 16.7 \\
\ce{^{52g}Mn_{(i)}} & 9   & 1.101 & 33.3  & 33.3 \\
\ce{^{58g}Co_{(i)}} & 9   & 0.472 & 66.7  & 55.6 \\
\ce{^{61}Zn_{(i)}}     & 7   & 0.944 & 100.0 & 100.0 \\
\ce{^{60}Cu_{(i)}}     & 6   & 0.768 & 100.0 & 100.0 \\
\ce{^{59}Fe_{(i)}}     & 5   & 0.709 & 80.0  & 80.0 \\
\ce{^{52m}Mn_{(i)}} & 1   & 0.854 & 100.0 & 0.0 \\
\ce{^{60}Zn_{(i)}}     & 1   & 0.439 & 100.0 & 100.0 \\
\ce{^{48}Sc_{(i)}}     & 1   & 0.516 & 100.0 & 100.0 \\
\ce{^{44m}Sc_{(i)}}     & 1   & 0.836 & 0.0   & 0.0 \\
\ce{^{44}Sc_{(i)}}     & 1   & 0.730 & 100.0 & 100.0 \Bstrut\\
\end{longtable}

The complete list of continuous parameter adjusted mean and standard deviation is provided in \autoref{tab:parameter_posteriors}

\begin{longtable}{lcccc}

\caption{Continuous parameter posterior statistics from iterative Bayesian updating. Default values are 1.0 for all adjustment parameters. Gen. 0 shows initial posterior after 3000 Sobol-sampled models; Gen. 5 shows final posterior after 5 refinement iterations (8000 total models). Mean and standard deviation reported as geometric statistics (mean $\mu_g$ and multiplicative factor $\sigma_{\times}$) due to lognormal prior distributions.}  \\   
\label{tab:parameter_posteriors}\\
\hline\hline
Parameter & Gen. 0 $\mu_g$ & Gen. 0 $\sigma_{\times}$ & Gen. 5 $\mu_g$ & Gen. 5 $\sigma_{\times}$ \\
\hline
\endfirsthead
\caption[]{(continued).}\\
\hline\hline
Parameter & Gen. 0 $\mu_g$ & Gen. 0 $\sigma_{\times}$ & Gen. 5 $\mu_g$ & Gen. 5 $\sigma_{\times}$ \\
\hline
\endhead
\hline \multicolumn{5}{r}{(Continued on next page)}\\
\endfoot

\hline\hline
\endlastfoot

$\sigma$ & 0.974 & 1.554 & 0.972 & 1.276 \Tstrut\\
$a_V^n$ & 0.953 & 1.611 & 0.981 & 1.266 \\
$a_V^p$ & 0.974 & 1.587 & 1.012 & 1.271 \\
$a_{V,so}^n$ & 1.001 & 1.656 & 0.987 & 1.282 \\
$a_{V,so}^p$ & 0.995 & 1.649 & 0.959 & 1.281 \\
$a_W^n$ & 0.995 & 1.654 & 0.987 & 1.299 \\
$a_W^p$ & 0.952 & 1.608 & 0.955 & 1.269 \\
$a_{W,so}^n$ & 1.003 & 1.663 & 0.981 & 1.292 \\
$a_{W,so}^p$ & 0.987 & 1.642 & 1.011 & 1.282 \\
$d_1^n$ & 0.997 & 1.648 & 1.026 & 1.283 \\
$d_1^p$ & 0.976 & 1.631 & 1.005 & 1.283 \\
$d_2^n$ & 0.987 & 1.631 & 0.977 & 1.275 \\
$d_2^p$ & 1.022 & 1.633 & 0.994 & 1.287 \\
$d_3^n$ & 1.004 & 1.635 & 0.996 & 1.276 \\
$d_3^p$ & 1.005 & 1.651 & 0.943 & 1.282 \\
$r_c^n$ & 0.987 & 1.658 & 0.980 & 1.281 \\
$r_c^p$ & 1.001 & 1.634 & 0.989 & 1.271 \\
$r_V^n$ & 0.955 & 1.609 & 0.949 & 1.268 \\
$r_V^p$ & 0.857 & 1.489 & 0.889 & 1.227 \\
$r_{V,so}^n$ & 1.003 & 1.637 & 0.971 & 1.290 \\
$r_{V,so}^p$ & 0.996 & 1.650 & 1.024 & 1.300 \\
$r_W^n$ & 0.990 & 1.632 & 1.025 & 1.284 \\
$r_W^p$ & 0.831 & 1.519 & 0.819 & 1.234 \\
$r_{W,so}^n$ & 0.997 & 1.667 & 0.991 & 1.288 \\
$r_{W,so}^p$ & 1.009 & 1.653 & 0.986 & 1.286 \\
$v_1^n$ & 0.983 & 1.647 & 1.001 & 1.283 \\
$v_1^p$ & 0.971 & 1.638 & 0.998 & 1.278 \\
$v_2^n$ & 0.992 & 1.646 & 0.975 & 1.295 \\
$v_2^p$ & 1.020 & 1.689 & 1.032 & 1.285 \\
$v_3^n$ & 0.994 & 1.652 & 1.024 & 1.294 \\
$v_3^p$ & 1.012 & 1.646 & 1.005 & 1.288 \\
$v_4^n$ & 0.997 & 1.668 & 0.966 & 1.277 \\
$v_4^p$ & 0.999 & 1.646 & 1.026 & 1.288 \\
$v_{so1}^n$ & 1.012 & 1.620 & 0.983 & 1.282 \\
$v_{so1}^p$ & 1.011 & 1.650 & 1.022 & 1.278 \\
$v_{so2}^n$ & 1.001 & 1.651 & 1.013 & 1.280 \\
$v_{so2}^p$ & 1.004 & 1.629 & 1.011 & 1.278 \\
$w_1^n$ & 0.987 & 1.665 & 1.023 & 1.291 \\
$w_1^p$ & 0.937 & 1.653 & 0.965 & 1.277 \\
$w_2^n$ & 0.984 & 1.644 & 1.007 & 1.288 \\
$w_2^p$ & 1.032 & 1.656 & 1.029 & 1.280 \\
$w_3^n$ & 1.008 & 1.646 & 0.984 & 1.281 \\
$w_3^p$ & 0.999 & 1.663 & 0.976 & 1.274 \\
$w_4^n$ & 1.001 & 1.639 & 1.035 & 1.286 \\
$w_4^p$ & 1.009 & 1.649 & 0.987 & 1.283 \\
$w_{so1}^n$ & 1.010 & 1.643 & 0.987 & 1.292 \\
$w_{so1}^p$ & 1.024 & 1.648 & 1.026 & 1.296 \\
$w_{so2}^n$ & 1.000 & 1.640 & 1.017 & 1.279 \\
$w_{so2}^p$ & 1.001 & 1.641 & 0.993 & 1.291 \\
$c_{\rm break}^{\alpha}$ & 1.002 & 1.656 & 0.993 & 1.284 \\
$c_{\rm break}^d$ & 0.993 & 1.646 & 1.046 & 1.277 \\
$c_{\rm break}^p$ & 0.995 & 1.648 & 1.022 & 1.281 \\
$c_{\rm break}^t$ & 1.004 & 1.662 & 0.971 & 1.296 \\
$c_{\rm knock}^{\alpha}$ & 1.011 & 1.640 & 1.016 & 1.284 \\
$c_{\rm knock}^d$ & 0.997 & 1.648 & 0.989 & 1.288 \\
$c_{\rm knock}^p$ & 1.012 & 1.647 & 0.997 & 1.304 \\
$c_{\rm knock}^t$ & 1.008 & 1.640 & 0.990 & 1.289 \\
$c_{\rm strip}^{\alpha}$ & 1.000 & 1.647 & 1.006 & 1.279 \\
$c_{\rm strip}^d$ & 0.997 & 1.651 & 1.018 & 1.272 \\
$c_{\rm strip}^p$ & 0.996 & 1.660 & 0.984 & 1.290 \\
$c_{\rm strip}^t$ & 1.007 & 1.648 & 1.029 & 1.290 \\
$C_1$ & 0.883 & 1.592 & 0.904 & 1.260 \\
$C_2$ & 1.006 & 1.648 & 0.962 & 1.296 \\
$C_3$ & 1.007 & 1.650 & 1.018 & 1.273 \\
$g$ & 0.993 & 1.633 & 1.036 & 1.293 \\
$r_{\nu\nu}$ & 0.953 & 1.675 & 0.977 & 1.291 \\
$r_{\nu\pi}$ & 0.976 & 1.651 & 0.934 & 1.268 \\
$r_{\pi\nu}$ & 0.985 & 1.650 & 0.997 & 1.290 \\
$r_{\pi\pi}$ & 1.006 & 1.663 & 0.968 & 1.283 \\
$\sigma_{\rm preeq}$ & 1.003 & 1.625 & 1.061 & 1.291 \Bstrut\\
\end{longtable}

\begin{figure*}

\subfloat[$r_w^p$]{\includegraphics[width=0.45\linewidth]{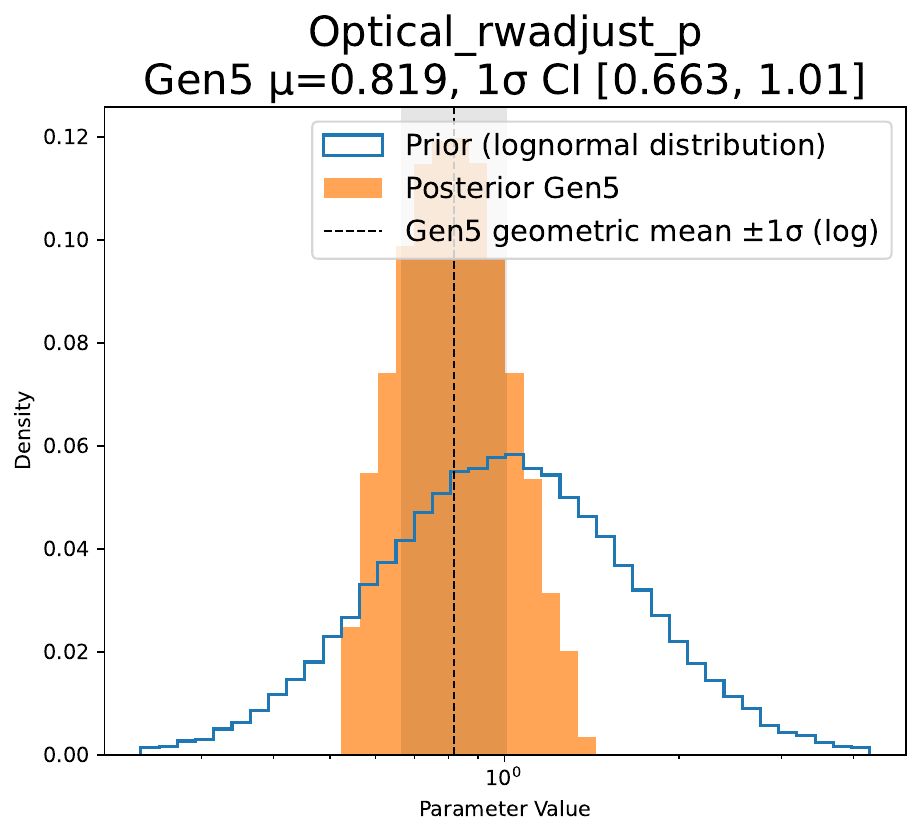}\label{fig:rwp}}
\subfloat[$r_v^p$]{\includegraphics[width=0.45\linewidth]{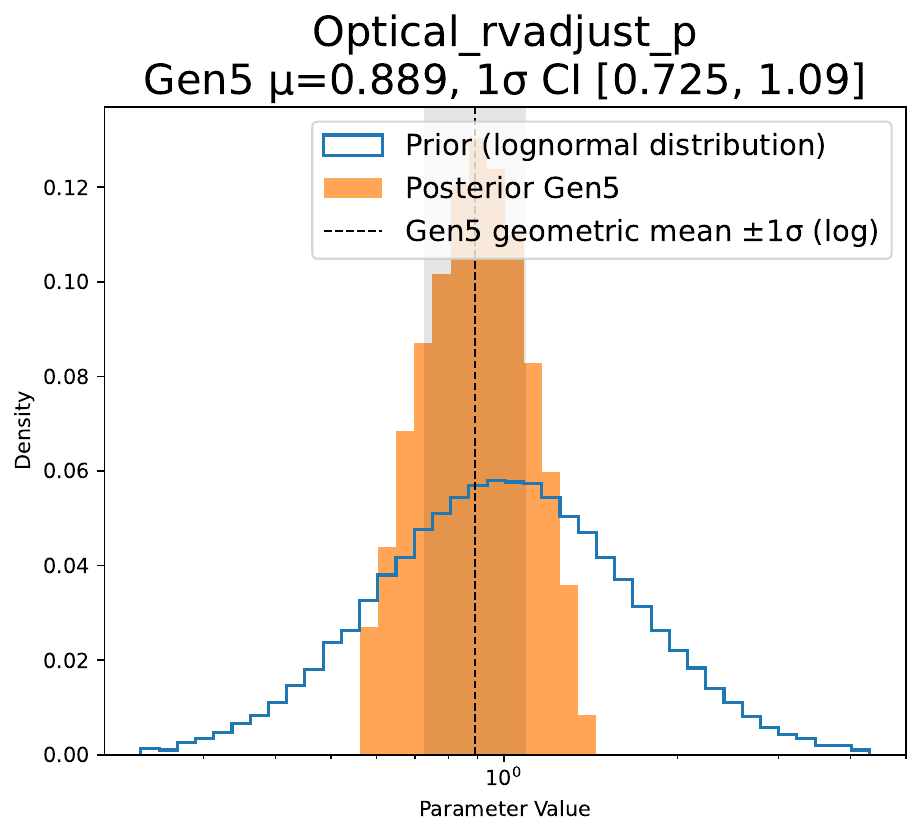}\label{fig:rvp}}\\
\subfloat[$r_v^n$]{\includegraphics[width=0.45\linewidth]{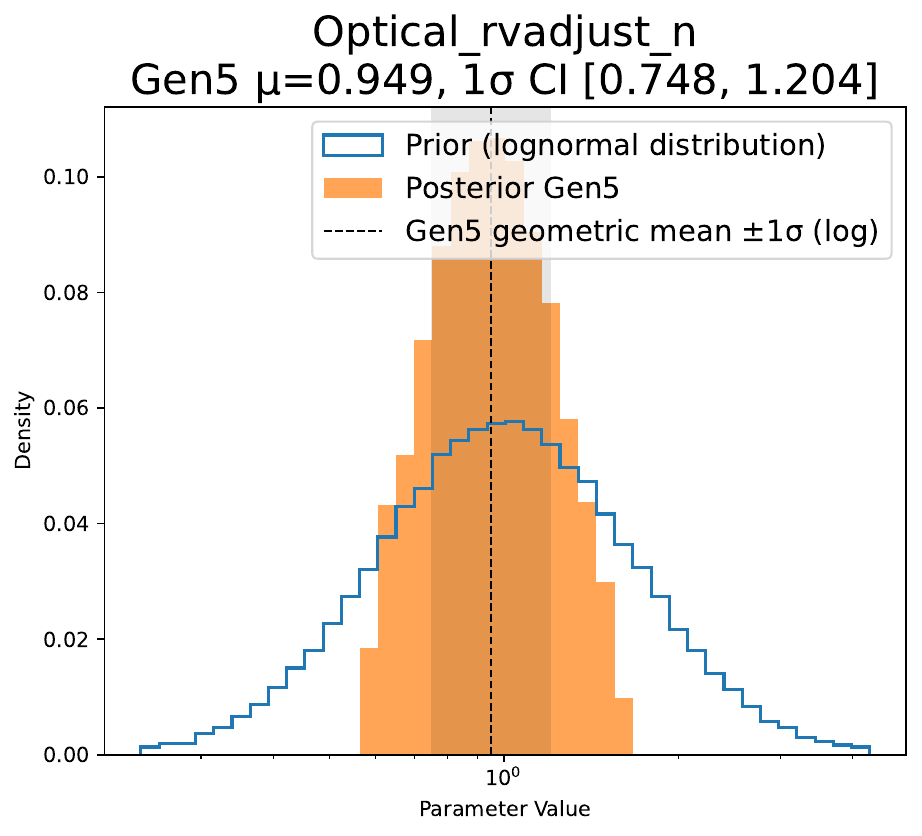}\label{fig:rvn}}
\subfloat[$D_3^p$]{\includegraphics[width=0.45\linewidth]{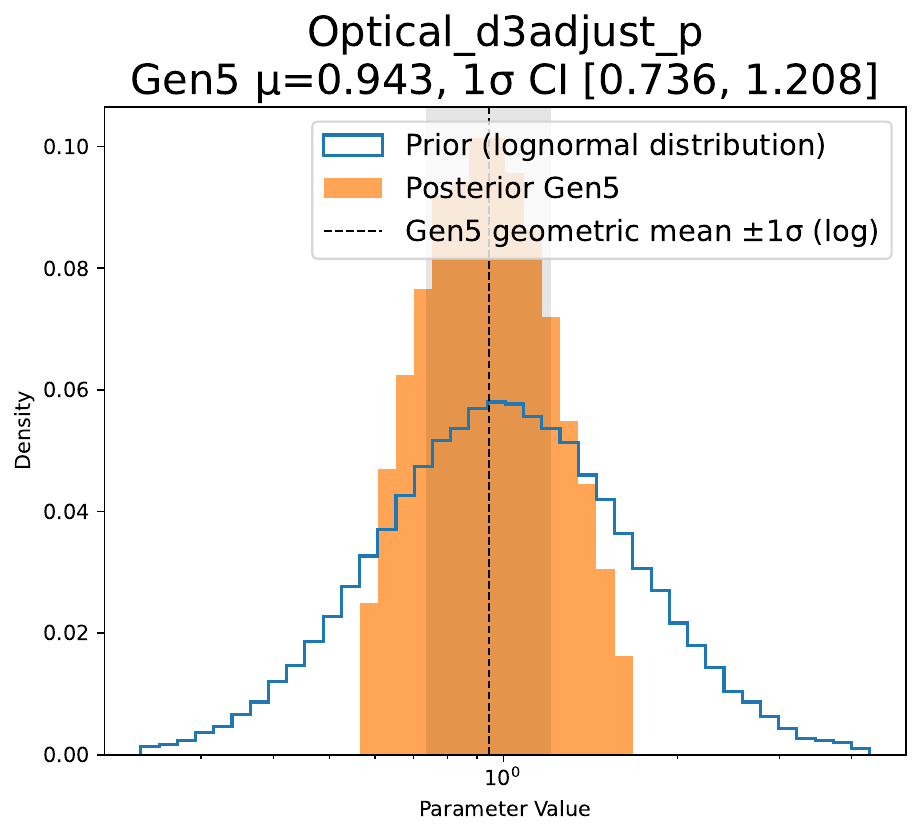}\label{fig:d3p}}
\caption{Optical model shifts of note during generational parameter permutations}
\end{figure*}

The majority of parameter adjustments made across the generations did not result in significant deviation from the TALYS default parameter values. Within the optical model parameters, $r_w^p$(\autoref{fig:rwp}), $r_v^p$(\autoref{fig:rvp}), $r_v^n$(\autoref{fig:rvn}) and $d_3^p$(\autoref{fig:d3p}) were the only parameters whose values by generation 5 had shifted by more than 5\%. However, it should be noted that $R_w^p$ shifted by over 18\%. 
\begin{figure*}

\subfloat[$C_1$]{\includegraphics[width=0.45\linewidth]{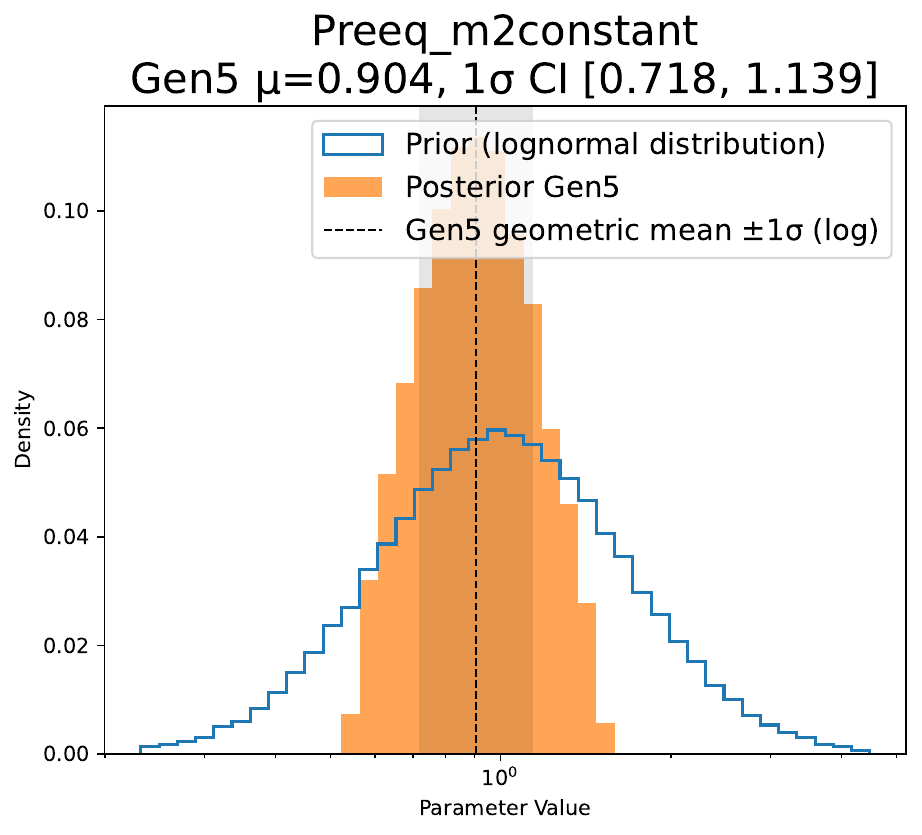}\label{fig:m2c}}
\subfloat[$r_{\nu\pi}$]{\includegraphics[width=0.45\linewidth]{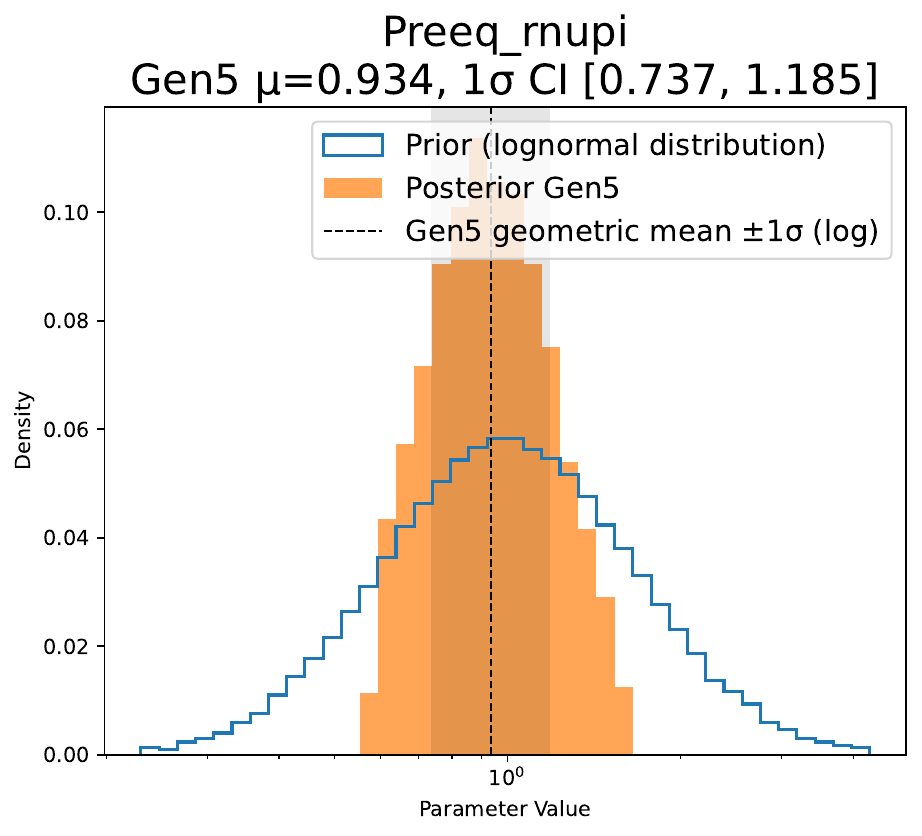}\label{fig:rnp}}\\
\caption{Pre-equilibrium model shifts during general parameter permutations}
\end{figure*}

For pre-equilibrium parameters that shifted the most were $C_1$ (\autoref{fig:m2c}) and $r_{\nu\pi}$ (\autoref{fig:rnp}), with the pre-equilibrium spin cut-off parameter $\sigma_{preeq}$ also shifting by greater than 5\%. This suggests that the TALYS default parameter values are well established for a more holistic approach for predicting cross-section values, with minor adjustments needed for specific parameters.

\subsection{Validation with Cumulative Cross Sections}

The initial 3000 inputs and final iteration (Generation 5) of 1000 TALYS runs was then compared separately to cumulative EXFOR data (MT=9001). This data refers to measurements of total production, which can include parent-feeding isotopes. The purpose of withholding this data from initial analysis was for validation of the results. This dataset consists of 23 isotopes, referenced in \autoref{tab:ind,cum} and over 1600 datapoints from 17 datasets \cite{gruetter_excitation_1982,voyles_excitation_2018,voyles_proton-induced_2021,orth_pion-induced_1978,garrido_new_2016,yashima_measurement_2003,heydegger_thin-target_1972,haller_relative_1961,williams_excitation_1967,greenwood_measurement_1984,lagarde-simonoff_spallation_1975,shahid_measurement_2015,fox_investigating_2021,uddin_experimental_2004,michel_cross_1997,graves_nuclear_2016,mills_experimental_1992}.
The overall performance from the initial generation to generation 5 provided some results in line with the independent cross section analysis, but differed in key areas.

\begin{figure}
    \centering
    \includegraphics[width=1\linewidth]{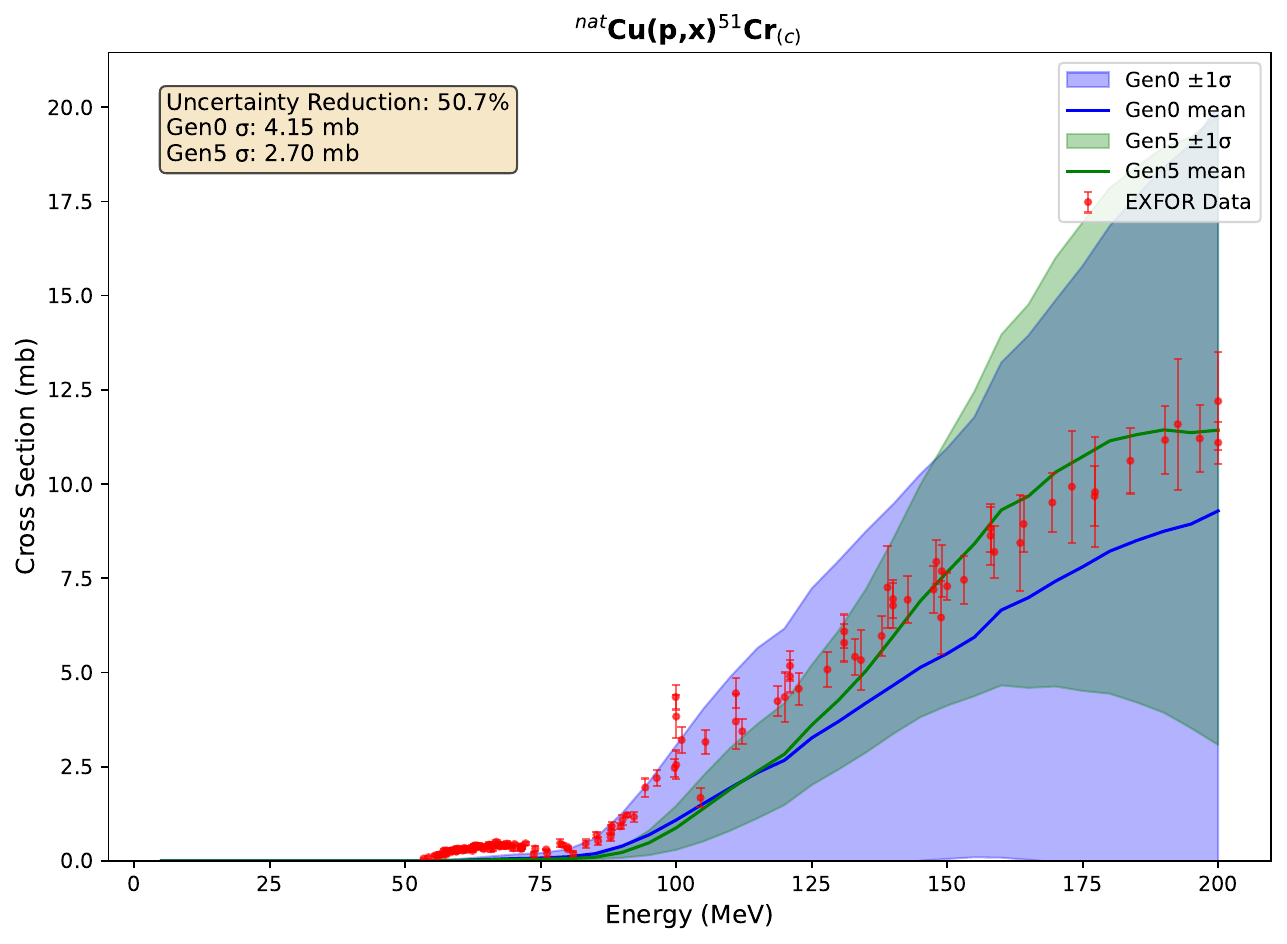}
    \caption{Visualization of the reduction in uncertainty band and improving prediction for the cumulative residual product, \ce{^{51}Cr}.}
    \label{fig:51Cr}
\end{figure}

\begin{figure}
    \centering
    \includegraphics[width=1\linewidth]{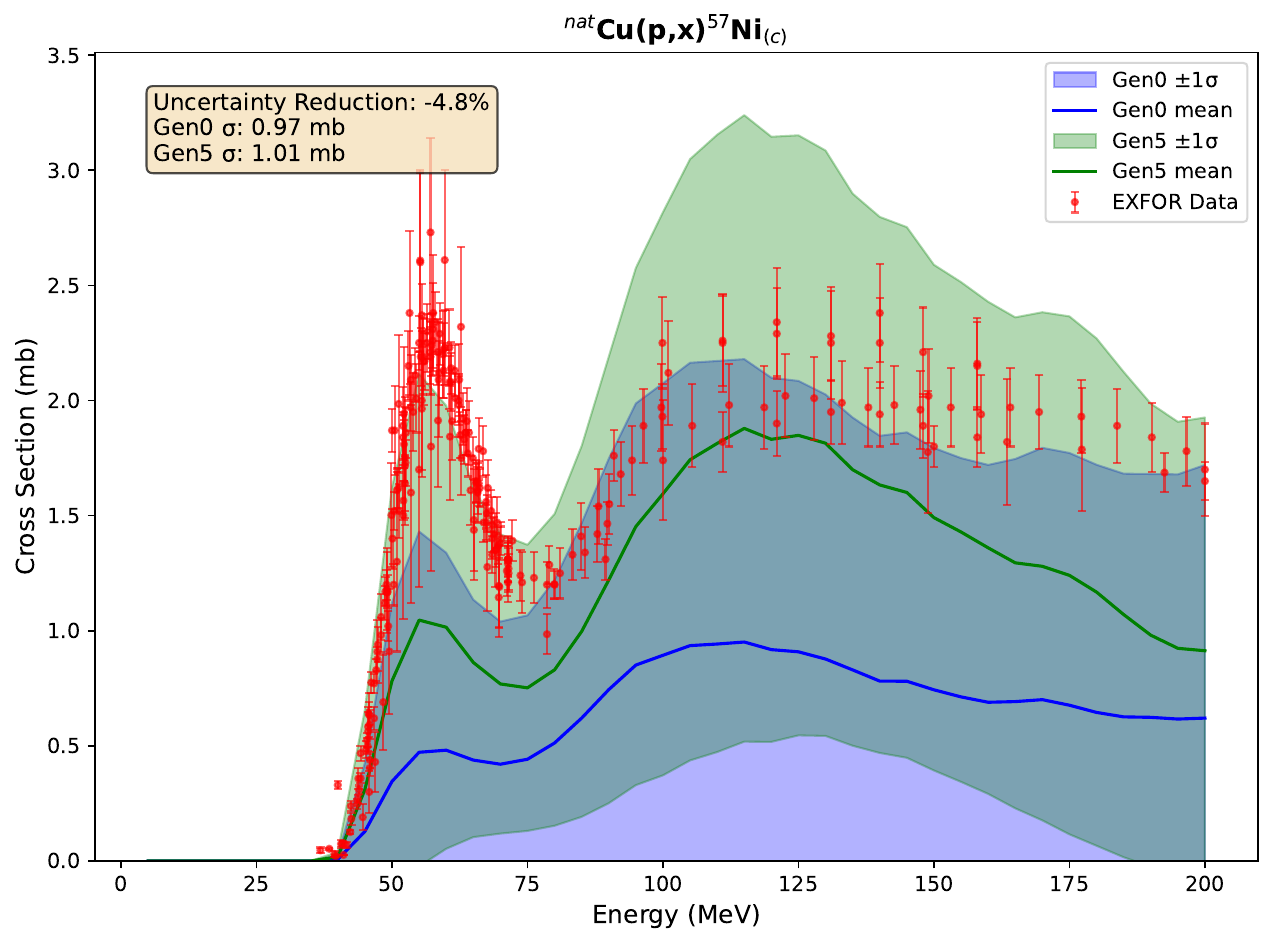}
    \caption{Visualization of the reduction in uncertainty band and improving prediction for the cumulative residual product, \ce{^{57}Ni}.}
    \label{fig:57Ni}
\end{figure}

Similar to the independent residual product cross sections, certain isotopes saw improvement in narrowing of uncertainty band, seen in \ce{^{51}Cr} in \autoref{fig:51Cr}, while others saw very little, or even worsening in the uncertainty band, as seen in \ce{^{57}Ni} in \autoref{fig:57Ni}. 
Overall performance for the 23 isotopes is available in \autoref{tab:cumulative_validation}.

\begin{table}[h]
\centering
\caption{Cumulative cross section validation results comparing Gen0 (initial posterior) and Gen5 (refined posterior). Columns show isotope, number of EXFOR data points, uncertainty reduction from Gen0 to Gen5, and containment (fraction of EXFOR points within $\pm 1\sigma$ posterior bands).}
\label{tab:cumulative_validation}
\begin{tabular}{lcccc}
\hline\hline
Isotope & N$_{\rm pts}$ & $\sigma_{\rm Gen5}/\sigma_{\rm Gen0}$ & Cont.$_{\rm Gen0}$ (\%) & Cont.$_{\rm Gen5}$ (\%) \\
\hline
\ce{^{57}Ni_{(c)}}    & 242 & 1.048 & 33.5  & 83.5 \\
\ce{^{61}Cu_{(c)}}    & 228 & 0.339 & 91.7  & 75.0 \\
\ce{^{59}Fe_{(c)}}    & 222 & 0.527 & 99.5  & 99.5 \\
\ce{^{51}Cr_{(c)}}    & 139 & 0.493 & 43.2  & 26.6 \\
\ce{^{56}Ni_{(c)}}    & 138 & 0.938 & 95.7  & 97.1 \\
\ce{^{57}Co_{(c)}}    & 117 & 0.614 & 31.6  & 11.1 \\
\ce{^{55}Co_{(c)}}    & 102 & 0.999 & 77.5  & 92.2 \\
\ce{^{60}Cu_{(c)}}    & 82  & 0.540 & 87.8  & 84.1 \\
\ce{^{52g}Mn_{(c)}} & 82  & 0.948 & 32.9  & 42.7 \\
\ce{^{56}Co_{(c)}}    & 69  & 0.681 & 46.4  & 27.5 \\
\ce{^{56}Mn_{(c)}}    & 69  & 0.439 & 89.9  & 62.3 \\
\ce{^{48}V_{(c)}}     & 48  & 0.953 & 72.9  & 89.6 \\
\ce{^{48}Cr_{(c)}}    & 25  & 1.077 & 92.0  & 52.0 \\
\ce{^{60}Co_{(c)}}    & 21  & 0.306 & 71.4  & 42.9 \\
\ce{^{52}Mn_{(c)}}    & 18  & 0.896 & 55.6  & 55.6 \\
\ce{^{52g}Fe_{(c)}} & 17  & 0.793 & 100.0 & 100.0 \\
\ce{^{47}Sc_{(c)}}    & 12  & 0.344 & 50.0  & 41.7 \\
\ce{^{61}Co_{(c)}}    & 9   & 0.172 & 33.3  & 0.0 \\
\ce{^{49}Cr_{(c)}}    & 7   & 0.780 & 100.0 & 0.0 \\
\ce{^{52}Fe_{(c)}}    & 4   & 0.727 & 100.0 & 100.0 \\
\ce{^{52m}Mn_{(c)}} & 4   & 0.744 & 0.0   & 0.0 \\
\ce{^{48}Sc_{(c)}}    & 2   & 0.415 & 100.0 & 100.0 \\
\ce{^{44}Sc_{(c)}}    & 2   & 0.399 & 100.0 & 50.0 \\
\hline\hline
\end{tabular}
\end{table}






In a broader view, it was observed that in contrast to the reduction in the median $\chi^2$ in the independent work of 71.7\%, and median $\chi^2$ of cumulative validation of 35.2\% the overall mean $\chi^2$ for the cumulative validation actually increased by a modest 6.7\% from the initial to final generation. 

Another key takeaway is the simultaneous reduction in the uncertainty bands, with an average reduction of 34.0\%, compared to the 32.2\% from the independent data, as well as the experimental data containment, which remained constant at ~68\%, again in line with the 66.6\% from independent data, suggesting that the uncertainty bands remained well calibrated for the validation channels.

This can be interpreted as a successfully narrowed posterior distribution and maintained uncertainty, but the narrowed parameter posterior distributions, centered on values that fit the independent reaction channels lose their fidelity when residual products cannot be disentangled from their decay chains, even when examining the same target isotope and incident particle. While these datasets were held out for validation purposes, future work could explore using all channels, both independent and cumulative, from the start of the iterative process. However, this could present other problems in analysis and interpretation.
\subsection{Investigation into a surrogate model}
With this large dataset, we conducted an investigation into whether machine learning could accelerate evaluations by approximating $\chi^2$ for TALYS inputs and automate the process of iterating through generations. 
A variety of models were tested, including neural networks, tree-based models, and Gaussian process-styled approaches. 
Since the initial generation included a variation in both models and parameters, it was determined that the best training dataset would be Generation 1, at which point the discrete models were frozen and only parameters were varied.
Additionally, the method of resampling in this iterative process limits the exploration beyond the value range established by the previous generation for each parameter, so it was predicted that the generated posterior distribution would become more concentrated, with ESS decreasing and weights concentrating on fewer samples.

In an attempt to design a surrogate, however, it was observed that ESS increased in size over each subsequent generation, with weights becoming more uniform and, most surprisingly, the reduction in uncertainty over each generation became negative, suggesting the posterior distribution was broadening.

These findings suggest that a surrogate-driven approach alone is not sufficient for the high 70-dimensional, small 1000 input sample dataset for training an accurate model. Significantly increasing the input size of properly calculated TALYS runs does not provide computational benefit. 

Nevertheless, despite the failure of predicting accurate $\chi^2$ values, some promising results were seen using a neural network in its ability to rank an input (i.e. which inputs would likely have a lower $\chi^2$ compared to one another). The degradation of accurate predictions, but relative maintenance of ranking, across the surrogate generations can be seen in \autoref{tab:generation_performance}.
This suggests the potential for a hybrid approach, where a surrogate could be used to downselect tens of thousands of generated inputs to the most likely best performers for each generation. 
This screening process can still provide benefit, potentially preventing the ESS collapse from occurring too rapidly, which here can be seen in \autoref{tab:generation_summary}.
Further research into this potential use will be explored in future work.

\begin{table}[t]
\caption{
Model performance by generation.}
\label{tab:generation_performance}
\begin{ruledtabular}
\begin{tabular}{lcc}
Generation & $R^2$ & Spearman $\rho$ \\
\hline
Gen2 & 0.716 & 0.879 \\
Gen3 & 0.603 & 0.844 \\
Gen4 & 0.343 & 0.790 \\
Gen5 & -0.433 & 0.770 \\
\end{tabular}
\end{ruledtabular}
\end{table}


        
    \section{\label{sec:Conclusions}Conclusions\protect}
        The work presented in this paper presents a systematic, iterative Bayesian Monte Carlo effort to establish uncertainty bands for nuclear reaction cross sections and parameters in the TALYS 2.0 nuclear reaction modeling code for \ce{^{nat}Cu}(p,x) reactions. The iterative sampling utilized 24 independent residual product cross section measurements from EXFOR. Substantial improvement was achieved from an initial, uninformed ensemble of TALYS model and parameter variations after five iterative generations. 

A reduction in the median $\chi^2$ was observed to decrease by 71.7\%. This demonstrates this iterative process was successful in adjusting the default value and uncertainties of continuous parameters in TALYS. 
Each successive generation, while reducing effective sample size, maintained significant coverage.
This suggests that the proposal distributions, truncated lognormal kernel density estimations, informed by the weighted posterior, were effective in reducing uncertainty distribution of parameter values. 
This was backed up by maintaining sufficient experimental data coverage within 1-$\sigma$ uncertainty bands of the posterior distribution.

Of note is the very small improvement in the best $\chi^2$ over generations, suggesting that the initial sampling of 3000 Sobol-sampled inputs had already explored the near optimal set of model/parameter combinations. 
This is not surprising, but it is not the purpose of this work, and subsequent generations served to refine parameters and reduce uncertainty. 
This further suggests a well-defined prior distribution, with a truncated lognormal distribution around physically motivated default TALYS values.



Further work, with significantly greater compute, could explore many more model combinations in the initial prior. This work does not suggest that the model combination chosen is perfect, but rather the best observed with available computational resources.

The use of held-out cumulative cross section experimental data served as a check on the results, but this check indicated that the improvements did not transfer to these other reactions, despite the same incident particle and target material. This illustrates the limitations from a holistic approach and highlights the need for physics-informed evaluations in parameter adjustments for uncertainty reduction in theoretical nuclear models.

A subsequent investigation into utilization of a surrogate model showed that while a simple neural network cannot accurately predict $\chi^2$ values and automatically iterate across generations, it can perform reasonably well in ranking input performance, suggesting the opportunity for a hybrid approach of surrogate-assisted down-selection and screening for the inputs likely to perform better than others.

Overall, while not intended as a replacement for evaluated nuclear data, this work presents an exploration of iterative Bayesian strategies that highlight both the strengths and weaknesses of purely theoretical charged-particle reaction modeling.

Data from this evaluation is available from the author or directly from Zenodo\cite{apgar_2026_23674681}.

    
    \begin{acknowledgments}
        This material is based upon work supported by the U.S. Department of Energy, Office of Science, Office of Isotope R\&D and Production, carried out under Los Alamos National Laboratory (Contract No. 89233218CNA000001), Lawrence Berkeley National Laboratory (Contract No. DE-AC02-05CH11231), Texas A\&M University Cyclotron Institute (Contract No. DE-FG02-93ER40773) and Brookhaven National Laboratory (Contract No. DEAC02-98CH10886), as well as the Department of Energy, National Nuclear Security Administration, under Award Number DE-NA0004150, the Center for Excellence in Nuclear Training And University-based Research (CENTAUR).

    \end{acknowledgments}

\appendix

\bibliography{references}

\end{document}